\documentclass[showpacs,aps,prd,preprint,nofootinbib,showkeys,unsortedaddress,raggedbottom]{revtex4-1}
\pdfoutput=1

\usepackage{bm}
\usepackage{amsmath}
\usepackage{amsfonts} 
\usepackage{graphicx}
\usepackage{subfigure}
\usepackage[usenames,dvipsnames]{color}
\definecolor{darkblue}{RGB}{0,0,196}
\usepackage[colorlinks=true,linkcolor=darkblue,citecolor=darkblue,urlcolor=darkblue]{hyperref}

\usepackage{setspace}
\usepackage{footmisc}
\usepackage{multirow}
\usepackage{hhline}

\def\be{\begin{equation}}
\def\ee{\end{equation}}
\def\ba{\begin{eqnarray}}
\def\ea{\end{eqnarray}}
\def\bew{\begin{equation*}}
\def\eew{\end{equation*}}

\begin{document}

\title{Heavy quarkonium suppression beyond the adiabatic limit}

\author{Jacob Boyd}
\email{jboyd29@kent.edu}
\author{Thomas Cook} 
\email{tcook22@kent.edu}
\author{Ajaharul Islam} 
\email{aislam2@kent.edu}
\author{Michael Strickland}
\email{mstrick6@kent.edu} 

\affiliation{Department of Physics, Kent State University, Kent, OH 44242 United States}

\begin{abstract}
Many prior studies of in-medium quarkonium suppression have implicitly made use of an adiabatic approximation in which it was assumed that the heavy quark potential is a slowly varying function of time.  In the adiabatic limit, one can separately determine the in-medium breakup rate and the medium time evolution, folding these together only at the end of the calculation.  In this paper, we relax this assumption by solving the 3d Schr\"odinger equation in real-time in order to compute quarkonium suppression dynamically.  We compare results obtained using the adiabatic approximation with real-time calculations for both harmonic oscillator and realistic complex heavy quark potentials. Using the latter, we find that, for the $\Upsilon(1s)$, the difference between the adiabatic approximation and full real-time evolution is at the few percent level, however, for the $\Upsilon(2s)$, we find that the correction can be as large as 18\% in low temperature regions.  For the $J/\Psi$, we find a larger difference between the dynamical evolution and the adiabatic approximation, with the error reaching approximately 36\%.
\end{abstract}

\date{\today}


\maketitle

\section{Introduction}

The primary goal of the ultrarelativistic heavy-ion collision (URHIC) program ongoing at Brookhaven National Laboratory's (BNL) Relativistic Heavy Ion Collider (RHIC) and the European Organization for Nuclear Research's (CERN) Large Hadron Collider (LHC) is to produce and study the properties of a deconfined state of matter called the quark-gluon plasma (QGP).  In the highest energy collisions at RHIC and LHC one probes the region of the QCD phase diagram corresponding to low baryochemical potential and high temperature.  In this region of the phase diagram, the QGP is believed to be created at temperatures exceeding the pseudo-critical temperature $T_{\rm pc} \simeq 155$ MeV.  It is called the pseudo-critical temperature because, at low baryochemical potential, the transition from a hot hadron gas to the QGP has been shown to be a crossover transition which interpolates smoothly between the hadronic and QGP phases \cite{Bazavov:2013txa,Borsanyi:2016bzg}.  At temperatures just above $T_{\rm pc}$ one finds the rapid breakup of light quark bound states such as pions, kaons, etc., however, since they have much larger binding energies, heavy quark bound states can survive to much higher temperatures.  Based on current estimates, the $J/\Psi$ and $\Upsilon(1s)$ states survive to temperatures of approximately 300-400 MeV and 600-700 MeV, respectively \cite{Andronic:2015wma,Mocsy:2013syh,Brambilla:2010cs}.  

In order to model the evolution, mixing, and breakup of heavy quark bound states one can make use of the heavy-quark limit of quantum chromodynamics (QCD) in which case one can formulate the problem in terms of a non-relativistic potential (pNRQCD) \cite{PhysRevD.21.203,Lucha:1991vn,Brambilla:2004jw,Brambilla:2010xn}.  This underlying non-relativistic picture was used implicitly in the original papers of Karsch, Matsui, and Satz who made the first predictions that heavy-quarkonia would ``melt'' in the QGP and that the precise way in which they melted could provide insight into properties of the QGP such as its initial temperature, degree of transverse expansion, etc. \cite{Matsui:1986dk,Karsch:1987pv}.  One thing that the original papers missed, however, was the fact that, if one self-consistently computes the heavy-quark potential in the high-temperature limit, one finds that the potential has both real and imaginary parts \cite{Laine:2006ns,Dumitru:2007hy,Brambilla:2008cx,Burnier:2009yu,Dumitru:2009fy,Dumitru:2009ni,Margotta:2011ta,Guo:2018vwy}.  In the high-temperature limit, the real part of the singlet potential reduces to a Debye-screened Coulomb potential, but the imaginary part has no classical analogue.  Physically, the imaginary part of the potential encodes the break-up rate of heavy-quarkonium bound states and can be properly understood in the context of open quantum systems in which the heavy-quark system is quantum mechanically coupled to a thermal heat bath \cite{Akamatsu:2011se,Akamatsu:2012vt,Akamatsu:2014qsa,Katz:2015qja,Brambilla:2016wgg,Kajimoto:2017rel,Brambilla:2017zei,Blaizot:2017ypk,Blaizot:2018oev,Yao:2018nmy}.

Soon after the realization that there was an imaginary part of the heavy-quark potential, the theoretical framework was applied to phenomenology \cite{Dumitru:2009ni,Margotta:2011ta,Strickland:2011mw,Strickland:2011aa,Strickland:2012cq,Krouppa:2015yoa,Krouppa:2016jcl,Krouppa:2017jlg}, however, in these prior works the authors made implicit use of the adiabatic approximation in which it was assumed that the time-scale for variation of the external parameters, e.g. temperature, was long compared to time-scale for the internal quantum dynamics.  In this limit, there is no mixing between the different quantum states and the final survival probability depends only on the imaginary part of a given state's eigen-energy integrated over the proper-time evolution of the QGP. It has been suggested in the literature, however, that there may be important corrections to results obtained in the adiabatic limit \cite{Dutta:2012nw}.

In this paper, we will make our first steps to go beyond the adiabatic approximation by numerically solving the Schr\"odinger equation in real-time using an efficient split-operator method.  We then construct the time-dependent eigenstates and compute the overlap of these with the time-evolved wave function.  We then compare our numerical results with simple formulae which result in the adiabatic limit.  We perform these tests using a toy complex harmonic oscillator potential and a realistic complex screened Coulomb potential.  In the latter case, we consider both bottomonium and charmonium bound states for a set of initial temperatures which are typically generated in URHICs and we use the same internal-energy based complex potential as has been used for prior phenomenological bottomonium suppression studies.  We find that, for the $\Upsilon(1s)$, the difference between the adiabatic approximation and full real-time evolution is at the few percent level, however, for the $\Upsilon(2s)$, we find that the correction can be as large as 18\% in low temperature regions of the QGP.  For the $J/\Psi$, we find a larger difference between the dynamical evolution and the adiabatic approximation, with the error reaching approximately 36\%.  These results indicate that a more quantitatively accurate description of heavy quarkonium suppression in the QGP requires solving the real-time Schr\"odinger equation coupled to a realistic 3d hydrodynamical background.

The structure of our paper is as follows.  In Sec.~\ref{sec:cho} we introduce the complex harmonic oscillator potential that we will use as a toy model.   In Sec.~\ref{sec:ckms} we introduce the complex version of the Karsch-Mehr-Satz potential which we will use to describe both charmonia and bottomonia.  In Sec.~\ref{sec:numerics} we discuss the numerical methods we will use to generate our results including a description of the real-time evolution algorithm and the algorithm for finding the time-dependent basis states necessary to project out the various eigenstates.  In Sec.~\ref{sec:results} we present our numerical results and comparisons with the adiabatic approximation in a set of illustrative cases.  Finally, in Sec.~\ref{sec:conclusions} we present our conclusions and an outlook for the future.

\section*{Notation and conventions}

We use natural units with $\hbar=c=k_B=1$.  Times and distances will be specified in GeV$^{-1}$.  Momenta and energies will be specified in GeV or MeV.

\section{One-dimensional complex harmonic oscillator}
\label{sec:cho}

As a simple toy model, we will consider a time-dependent complex harmonic oscillator (CHO) potential.  At a given time $t$ the potential is of the form
\be
V(x,t) = \frac{1}{2} k(t) x^2 \, ,
\label{eq:cho}
\ee
where $k(t)$ is a complex-valued function.  In the case that $k$ is complex, the oscillation frequency $\omega = \sqrt{k/m}$ is also complex.  In order for a stable ground state to exist, one should have $\Re(\omega) \geq 0$ and $\Im(\omega) \leq 0$ \cite{Margotta:2011ta}.  For any $k$ which satisfies these constraints, one can obtain the eigenfunctions and energies by analytic continuation of the standard harmonic oscillator (HO) basis functions (See Appendix of Ref.~\cite{Margotta:2011ta})\footnote{In practice, we only use the time-dependent eigenstates of the Hamiltonian to project out the overlaps at a given time.  This is done to be completely compatible with the assumptions made in the adiabatic approximation, in which case one assumes that at each time $t$ one can find the instantaneous change in the survival probability by solving the time-independent Schr\"odinger equation (SE) (see e.g. Refs.~\cite{Strickland:2011aa,Krouppa:2016jcl,Krouppa:2015yoa,Krouppa:2017jlg}).  The use of these time-dependent eigenstates allows us to make an apples-to-apples comparison with the adiabatic approximation.}
\be
\psi_n(t,x) = N_n e^{-\frac{1}{2} \alpha x^2} H_n(\sqrt{\alpha} x) \, ,
\label{eq:chobasis}
\ee
where $\alpha = m \omega$ and $N_n$ is the normalization constant computed by demanding that \mbox{$\langle \psi_n | \psi_n \rangle = 1$}.  The energy eigenvalues are the same as in the case of real-valued $k$
\be
E_n = \omega \left( n + \frac{1}{2} \right) .
\label{eq:hoeigen}
\ee
Importantly, for complex $k$ one finds that the basis functions $\psi_n$ are not orthogonal, i.e. $\langle \psi_n | \psi_m \rangle \neq 0$ for $n \neq m$.  However, they do form a linearly independent basis that we can use to decompose any time-dependent wave function, i.e.
\be
\psi(t,x) = \sum_{n=0}^\infty c_n(t) \psi_n(t,x) \, .
\label{eq:basis1}
\ee

When the spring constant $k$ is complex, the Hamiltonian is not Hermitian ($H \neq H^\dagger$) and the time evolution is not unitary.  In addition, as mentioned above, the energy eigenstates are no longer orthogonal.  Since this is not typical, it is important to establish some basic relations in this case.
Using \eqref{eq:basis1} we can obtain
\begin{eqnarray}
\langle \psi_{m} | \psi \rangle &=& O_{mn}(t) \, c_{n}(t) \, ,
\end{eqnarray}
where the sum over $n$ is implied, $O_{mn}(t) \equiv \langle \psi_{m} | \psi_{n} \rangle$, and we have dropped the arguments $t$ and $x$ of $\psi$ and $\psi_n$ for compactness.  Note, importantly, that if $k(t)$ is time dependent, then the basis functions $\psi_n$ depend on time.

Since the states are linearly independent, the determinant of $O_{mn}$ is non-zero and we can invert the matrix $O_{mn}$ to obtain
\be
c_{n}(t) = O^{-1}_{nm}(t) \langle \psi_{m} | \psi \rangle \, .
\ee
Using this formula one can extract the quantum amplitude of the $n$-th time-dependent quantum state from the total wave function.  This procedure is general and will also be used for other complex potentials including the one presented in the next section.

\section{Debye-screened coulomb potential plus imaginary part}
\label{sec:ckms}

We will also present results using a potential model based on high-temperature quantum chromodynamics (QCD) which provides information about the short- and medium-range parts of the heavy-quark potential~\cite{Laine:2006ns}.  For describing finite-temperature states which can have large radii compared to the confinement scale $\Lambda_{\rm QCD}^{-1} \sim 300$ MeV $\sim 0.66$ fm/c, however, one must supplement the perturbative potential obtained in Ref.~\cite{Laine:2006ns}
by a long range contribution.  For this purpose, we will make use of a Karsch-Mehr-Satz type potential based on the internal energy~\cite{Karsch:1987pv,Dumitru:2009ni,Strickland:2011aa}
\begin{eqnarray} 
\label{eq:repot}
\Re[V] &=& -\frac{\alpha}{r} \left(1+m_D\, r\right) \exp\!\left( -m_D
\, r  \right) + \frac{2\sigma}{m_D}\left[1-\exp\!\left( -m_D
\, r  \right)\right] \nonumber \\
&& \hspace{2cm} - \sigma \,r\, \exp(-m_D\,r) - \frac{0.8 \, \sigma}{m_Q^2\, r} \, ,
\label{eq:ckmspot}
\end{eqnarray}
where the last term is a finite quark mass correction \cite{Bali:1997am}, $\sigma = 0.210\;{\rm GeV}^2$ is the string tension, and $m_D$ is the QGP Debye mass $m_D^2 = (1.4)^2 \cdot N_c (1+N_f/6)  \, 4  \pi \alpha_s  \, T^2/3$ with $N_c=3$ and $N_f=2$.  The factor of $(1.4)^2$ takes into account higher-order corrections and has been determined using lattice QCD simulations~\cite{Kaczmarek:2004gv}.  We fix the Coulomb interaction strength to $\alpha = 4 \alpha_s/3 = 0.385$ to match the low temperature binding energy and mass spectrum of heavy quark states \cite{Dumitru:2009ni}.

For the imaginary part of the model potential we use the result obtained from leading-order finite temperature perturbation theory~\cite{Laine:2006ns}
\begin{equation} 
\Im[V] = - \alpha_s T \phi(\hat{r}) \, ,
\label{eq:impot}
\end{equation}
where $\hat{r} \equiv m_D r$, $\alpha_s = 0.289$, and
\begin{equation}
 \phi(\hat{r}) \equiv 2\int_0^{\infty}dz \frac{
z}{(z^2+1)^2} \left[1-\frac{\sin(z\, \hat{r})}{z\, \hat{r}}\right] .
\end{equation}

The total complex Karsch-Mehr-Satz (CKMS) potential is a sum of the real \eqref{eq:ckmspot} and imaginary \eqref{eq:impot} parts\,\footnote{This potential has also been referred to as the Strickland-Bazow potential in the literature since it was used in Ref.~\cite{Strickland:2011aa}.}
\be
V(r,t) = \Re[V] + i \, \Im[V] \, ,
\label{eq:ckmsdef}
\ee
where the time-dependence of the potential results from the fact that the temperature and hence the Debye mass $m_D$ depend on the proper-time.  All other parameters are held constant.

\section{Numerical algorithm for solving the time-dependent\\Schr\"odinger equation}
\label{sec:numerics}

Our goal is to solve the time-dependent Schr\"odinger equation with both the complex one-dimensional harmonic oscillator (CHO) \eqref{eq:cho} and CKMS potentials \eqref{eq:ckmsdef}.  The numerical algorithm will be slightly different in each case since the CHO wave function has support in $-\infty < x < \infty$ while the CKMS wave function has support only in $0 \leq r < \infty$ and must vanish at $r=0$.  We will start by describing the algorithm for the case of the CHO.

\subsection{CHO algorithm}

To begin, we discretize space using a lattice covering $-L/2 \leq x \leq L/2$ with a step size $\Delta x = L/N$ where $N$ is the number of points used to describe the wave function.  We also introduce a temporal grid with spacing $\Delta t$.  The spatial grid points in this case are given by $x_i = -L/2 + (i-1)\Delta x$ with $i \in \{1 \cdots N \}$.

As is well-known, the time evolution of the quantum mechanical wave function can be obtained in general using the time-evolution operator
\be
\psi(x,t+\Delta t) = \hat{U}_\Delta(t) \psi(x,t) \, ,
\ee
where
\be
\hat{U}_\Delta(t) = \exp(-i \hat{H}(t) \Delta t) \, ,
\ee
is the infinitesimal time-evolution operator.  To proceed, we can use the fact that $\Delta t$ is small to approximate\,\footnote{This formula and the scaling of the error due to a finite time step can be derived using the Baker-Campbell-Hausdorff theorem.}
%
\be
\exp(-i \hat{H}(t) \Delta t) \simeq \exp(- i V(x,t) \Delta t/2) \exp\!\left(-i \frac{\hat{p}^2}{2 m} \Delta t\right) \exp(- i V(x,t) \Delta t/2) + {\cal O}( (\Delta t)^2 ) \, ,
\ee
%
where we have factorized the time-evolution operator into two terms (first and last) which take care of the evolution due to the potential and a separate term (middle) which takes care of the evolution due to the kinetic energy.  

To make a single time step, we begin in configuration space with the initial wave function $\psi_0(x,t)$ and apply the rightmost exponential to obtain
\be
\psi_1(x,t) = \exp(- i V(x,t) \Delta t/2) \psi_0(x,t) \, .
\ee
For the momentum update, it is most convenient to work in momentum space, so we take the discrete Fourier transform of $\psi_1(x,t)$ to obtain the complex Fourier transform
\be
\tilde\psi_1(p,t) = \mathbb{F}[\psi_1(x,t)] \, ,
\ee
where $p$ are the discrete momenta associated with the spatial grid, $p_i = (i-N/2)\Delta k$ with $\Delta k = 2\pi/L$ and $i \in \{1 \cdots N \}$.  After transforming to momentum space one can apply the kinetic energy update directly
\be
\tilde\psi_2(p,t) =  \exp\!\left(-i \frac{p^2}{2 m} \Delta t\right) \tilde\psi_1(p,t) \, .
\ee
To make the final step of the evolution we return to configuration space using the inverse discrete Fourier transform
\be
\psi_2(x,t) =  \mathbb{F}^{-1}[\tilde\psi_2(p,t)] \, ,
\ee
followed by
\be
\psi_3(x,t) =  \exp(- i V(x,t) \Delta t/2) \psi_2(x,t) \, .
\ee
The result stored in $\psi_3(x,t)$ contains the updated wave function, i.e. $\psi(x,t + \Delta t) = \psi_3(x,t)$.  By repeating this procedure, we can evolve the wave function forward in time in a manifestly unitary manner which is generally faster and more accurate than traditional discrete evolution equations that work solely in configuration space.\footnote{We will work with complex potentials so the evolution will be non-unitary in the end; however, we want to make sure that any non-unitary behavior is physical in origin and not numerical.}$^,$\footnote{See App. \ref{app:a} for detailed benchmarks and discussion.}  This method is known in the literature as a split-step Fourier or pseudospectral method \cite{mclachlan_quispel_2002}.  Note that, although we assumed that the system was one-dimensional above, pseudospectral methods can be straightforwardly extended to an arbitrary number of spatial dimensions.

\subsection{CKMS algorithm}

The complex KMS potential is a purely radial potential which allows us to write the general solution in spherical coordinates as
\be
\psi(r,\theta,\phi,t) = \sum_{\ell m} R_{\ell m}(r,t) Y_{\ell m}(\theta,\phi) \, ,
\ee
where $Y_{\ell m}$ are spherical harmonics.  It is convenient to change variables to \mbox{$u_{\ell m}(r,t) \equiv r R_{\ell m}(r,t)$} to obtain 
\be
u(r,\theta,\phi,t) = \sum_{\ell m} u_{\ell m}(r,t) Y_{\ell m}(\theta,\phi) \, ,
\ee
where $u(r,\theta,\phi,t) = r \psi(r,\theta,\phi,t)$.  

This change of variables allows us to cast the Hamiltonian in one-dimensional form
\be
\hat{H}_\ell = \frac{\hat{p}^2}{2m} + V_{{\rm eff},\ell}(r,t) \, ,
\ee
where $V_{{\rm eff},\ell}(r,t) = V(r,t) + \frac{\ell(\ell+1)}{2 m r^2}$ and $\hat{p} = -i \frac{d}{dr}$.  At a fixed time $t$ the eigenstates satisfy
\be
\hat{H}_\ell u_\ell(r,t)  = E_\ell(t) u_\ell(r,t) \, ,
\label{eq:radialevaleq}
\ee
where we have used the fact that the eigenvalue equation \eqref{eq:radialevaleq} does not depend on the azimuthal quantum number $m$ to simplify $u_{\ell m} \rightarrow u_\ell$.   However, for the time evolution, we will keep the azimuthal quantum number for generality:
\be
u(r,\theta,\phi,t) = {\cal N} \sum_{\ell=\ell_{\rm min}}^{\ell_{\rm max}} \sum_{m=-\ell}^\ell \frac{1}{\sqrt{2\ell+1}} u_{\ell,m}(r,t) Y_{\ell m}(\theta,\phi) \, ,
\ee
where the factor of $(2\ell+1)^{-1/2}$ has been introduced by hand for simplifying the forthcoming normalization of $u$ and ${\cal N}$ is a normalization constant to be determined later.

Our goal is to obtain an evolution equation for each of the coefficient functions $u_{\ell,m}(r,t)$.  To proceed we use the orthogonality of the spherical harmonics, $\int d\Omega \, Y_{\ell' m'}^*(\theta,\phi) Y_{\ell m}(\theta,\phi) = \delta_{\ell \ell'} \delta_{m m'}$ where $d\Omega = d\phi \, d(\cos \theta)$ is the three-dimensional solid angle differential.  Using this, one obtains
\be
u_{\ell,m}(r,t) = \sqrt{2\ell+1} \int d\Omega \, Y_{\ell m}^*(\theta,\phi) \, u(r,\theta,\phi,t) \, .
\label{eq:u1}
\ee

As with the CHO potential we will use a pseudo-spectral method.  Applying the time evolution operator to $u$ gives
\ba
u(r,\theta,\phi,t + \Delta t) &=& \exp(- i \hat{H} \Delta t) u(r,\theta,\phi,t ) \nonumber \\
&=& {\cal N}  \sum_{\ell,m} \frac{1}{\sqrt{2\ell+1}} \exp(- i \hat{H} \Delta t) u_\ell(r,t) Y_{\ell m}(\theta,\phi) \nonumber \\
&=&  {\cal N} \sum_{\ell,m} \frac{1}{\sqrt{2\ell+1}} Y_{\ell m}(\theta,\phi) \exp(- i \hat{H}_\ell \Delta t) u_\ell(r,t)  \, ,
\ea
where the summation limits are implicit.  Using Eq.~\eqref{eq:u1}, one obtains
\be
u_{\ell,m}(r,t + \Delta t) = \exp(- i \hat{H}_\ell \Delta t) u_{\ell,m}(r,t) \, ,
\label{eq:uUpdate}
\ee
which tells us that we can update each of the different $\ell,m$ states independently using the corresponding Hamiltonian.  We note that the time evolution operator Eq.~\eqref{eq:uUpdate} is independent of $m$.

Finally, we need to come up with a prescription for the normalization of the various states.  Herein we take
\be
\int_0^\infty dr \, u_{\ell,m}^*(r,t) u_{\ell,m}(r,t) = 1 \, .
\ee
Using this and requiring that the total wave function is normalized
\be
\int dr d\Omega \, u^*(r,\theta,\phi,t) u(r,\theta,\phi,t) = 1 \, ,
\ee
gives ${\cal N} = 1/\sqrt{\ell_{\rm max} - \ell_{\rm min} + 1}$ and our final expression for the general wave function
\be
u(r,\theta,\phi,t) = \frac{1}{\sqrt{\ell_{\rm max} - \ell_{\rm min} + 1}} \sum_{\ell=\ell_{\rm min}}^{\ell_{\rm max}} \sum_{m=-\ell}^\ell \frac{1}{\sqrt{2\ell+1}} u_{\ell,m}(r,t) Y_{\ell m}(\theta,\phi) \, .
\ee

\subsubsection{Time evolution algorithm}

Similar to the CHO case, we will use the time evolution operator \eqref{eq:uUpdate} to evolve the wave function given a particular initial condition.  One complication compared to the CHO case is that the function $u_\ell(r,t)$ must vanish at the origin.  In order to enforce this we use real-valued Fourier sine series to describe both the real and imaginary parts of the wave function.  In this manner we guarantee that the correct boundary conditions at $r=0$ are satisfied automatically.  The resulting update steps are similar to those used for the CHO case but with $\mathbb{F} \rightarrow \mathbb{F}_s$:
\begin{enumerate}
\item Update in configuration space using a half-step: $ \psi_1 = \exp(- i V \Delta t/2) \psi_0$.
\item Perform Fourier sine transformations on real and imaginary parts separately:  \\ $\tilde\psi_1 = \mathbb{F}_s[\Re \psi_1] + i \mathbb{F}_s[\Im \psi_1] $.
\item Update in momentum space using: $\tilde\psi_2 =  \exp\!\left(-i \frac{p^2}{2 m} \Delta t\right) \tilde\psi_1$.
\item Perform inverse Fourier sine transformations on real and imaginary parts separately:  \\ $\psi_2 = \mathbb{F}_s^{-1}[\Re \tilde\psi_2] + i \mathbb{F}_s^{-1}[\Im \tilde\psi_2] $.
\item Update in configuration space using a half-step: $ \psi_3 = \exp(- i V \Delta t/2) \psi_2$.
\end{enumerate}

\subsubsection{Time-dependent basis functions}
\label{sec:cpas}

In the case of the CHO we could determine the quantum mechanical wave functions by simple analytical continuation of the HO wave function (see Eq.~\eqref{eq:chobasis}).  In the case of the CKMS potential it is not possible to determine the eigenfunctions analytically.  Instead, we compute them using a modified ``point and shoot'' method.  For the basis functions, at any time $t$, one must solve
\be
-\frac{1}{2m} \frac{d^2 u_\ell(r)}{dr^2} + V_{\rm eff,\ell}(r) u_\ell(r) = E u_\ell(r) \, ,
\label{eq:cpotu}
\ee
where both $E$ and $u_\ell$ are unknowns and there may be multiple solutions.  This equation must be solved subject to the boundary condition
\be
\lim_{r \rightarrow 0} u_\ell(r) = A r^{\ell +1} \, ,
\ee
where $A$ is an arbitrary constant which can be taken to be one.
For a given $\ell$ we start from a very small value of $r = r_{\rm min}$ and fix $u_\ell(r_{\rm min}) = r_{\rm min}^{\ell +1}$ and $u_\ell'(r_{\rm min}) = (\ell+1) r_{\rm min}^\ell$.  We then choose a trial value for $E$ and integrate Eq.~\eqref{eq:cpotu} from $r_{\rm min}$ to a large radius $r = r_{\rm max}$.  In order to be a normalizable solution, $u$ must vanish as $r \rightarrow \infty$.  For a real-valued potential, one can find all bound state solutions by making steps in E from a negative value to $E = V_{\rm max}$ and searching for the points where $u(r_{\rm max}) = 0$.  The set of values obtained in this manner, $E_n$, are the energy eigenvalues and the associated normalized solutions are the eigenfunctions $u_{n\ell}$.

In the case that the potential is complex, the energy eigenvalues become complex and the solutions using the point-and-shoot method become more difficult to obtain.  To surmount this difficulty we split the potential into real and imaginary parts and introduce an adjustable parameter $\delta$, i.e.
\be
V = \Re[V] + i \, \delta \, \Im[V] \, .
\ee
We then first find the solutions for $\delta = 0$ using the standard point-and-shoot method to identify the eigenvalues.  We then increment $\delta$ in steps of 0.1 until it takes the value $\delta =1$, using the previous step's solutions for the eigenvalues as the guess for the current step.  In this way we can find the complex energy eigenvalues and eigenfunctions efficiently.  These can then be saved to disk for later use.

\section{Results}
\label{sec:results}

We now turn to our numerical results.  In what follows we will make comparisons to the ``adiabatic approximation''.  In the adiabatic approximation, one ignores the mixing of various excited states which emerge in a time-dependent potential.  The result is that in this approximation the time-evolution of each state is governed solely by the time-dependent imaginary part of the states energy, i.e.
\be
N_n(t) = N_n(0) \exp\!\left( 2 \int_{t_0}^{t_f} dt \, \Im[E_n(t)] \right) ,
\label{eq:adiabaticapprx}
\ee
where $N_n(t) = |c_n(t)|^2$ is the modulus-squared overlap states of type $n$ at time $t$.  In this case, one can solve for the time-dependent energy eigenvalues numerically or analytically, e.g. for the CHO using Eq.~\eqref{eq:hoeigen} one has $\Im[E_n(t)] = (n+\frac{1}{2}) \Im[\omega(t)] =  (n+\frac{1}{2}) \Im[\sqrt{k(t)}]/m$.  This is expected to be a good approximation if the potential is changing slowly relative to the natural timescale of the quantum evolution.

\subsection{CHO potential results}

\begin{figure*}[t!]
\centerline{
\includegraphics[width=0.45\linewidth]{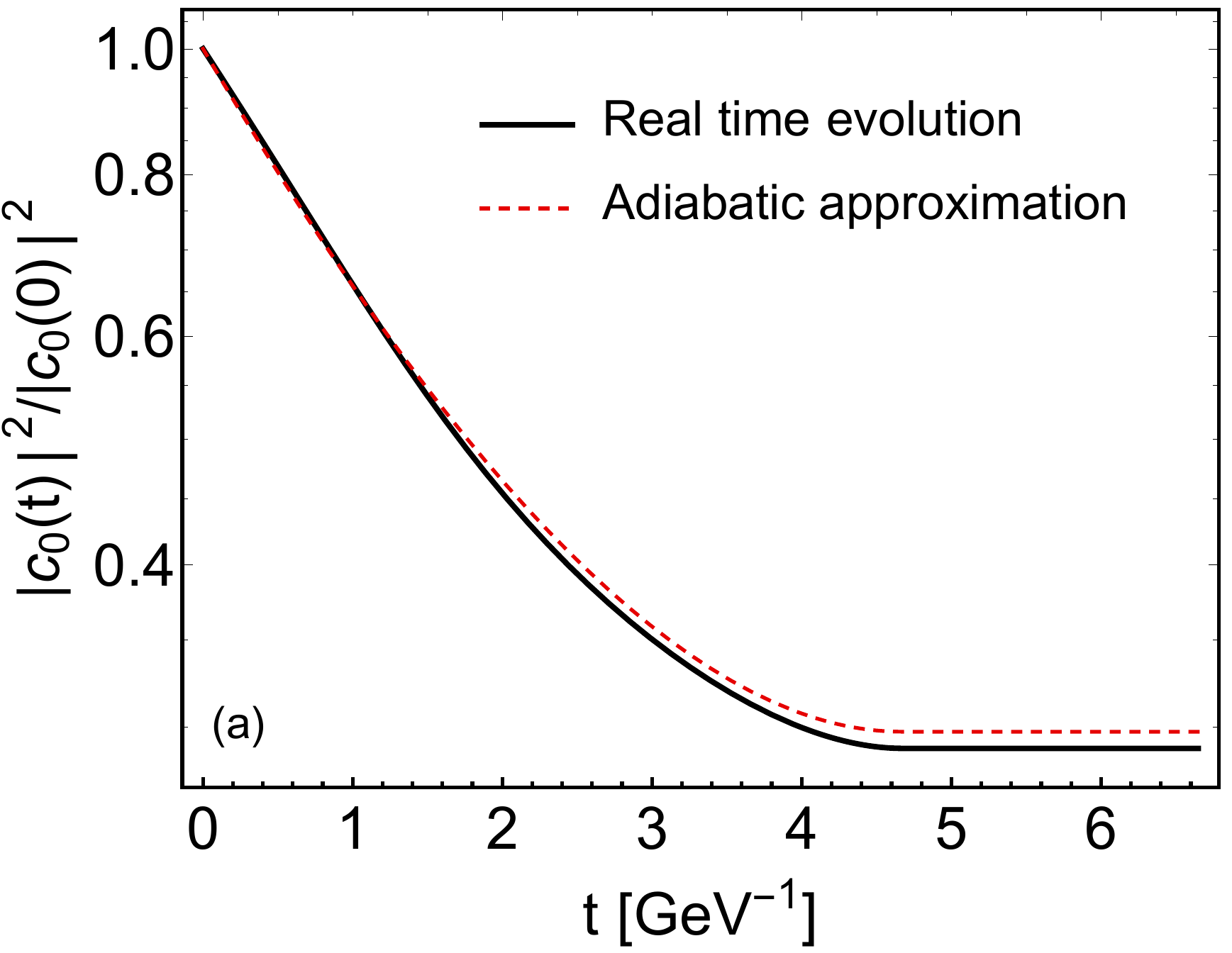}\hspace{5mm}
\includegraphics[width=0.48\linewidth]{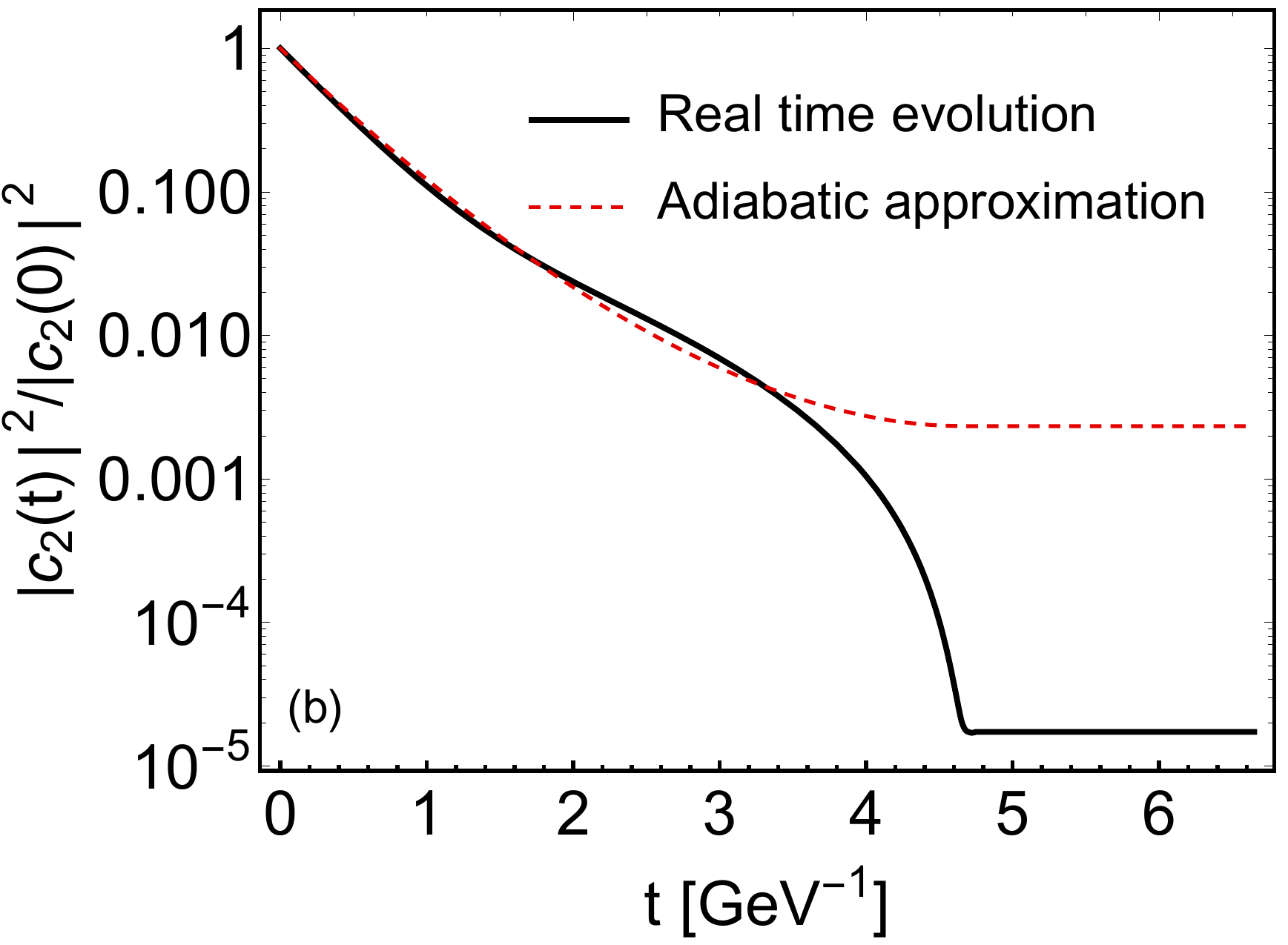}
}
\caption{Comparison of the adiabatic approximation and real-time evolution for the CHO potential with $t_2 = 4.67$ GeV$^{-1}$ in Eqs.~\eqref{eq:cho1} and \eqref{eq:cho2}.  Panel (a) shows the ground state ($n=0$) overlap coefficient squared normalized to its initial value and panel (b) shows the second excited state ($n=2$) overlap coefficient squared normalized to its initial value.  The solid black line is the real-time solution and the dashed red line is the adiabatic approximation.
}
\label{fig:cs}
\end{figure*}

\begin{figure*}[t!]
\centerline{
\includegraphics[width=0.47\linewidth]{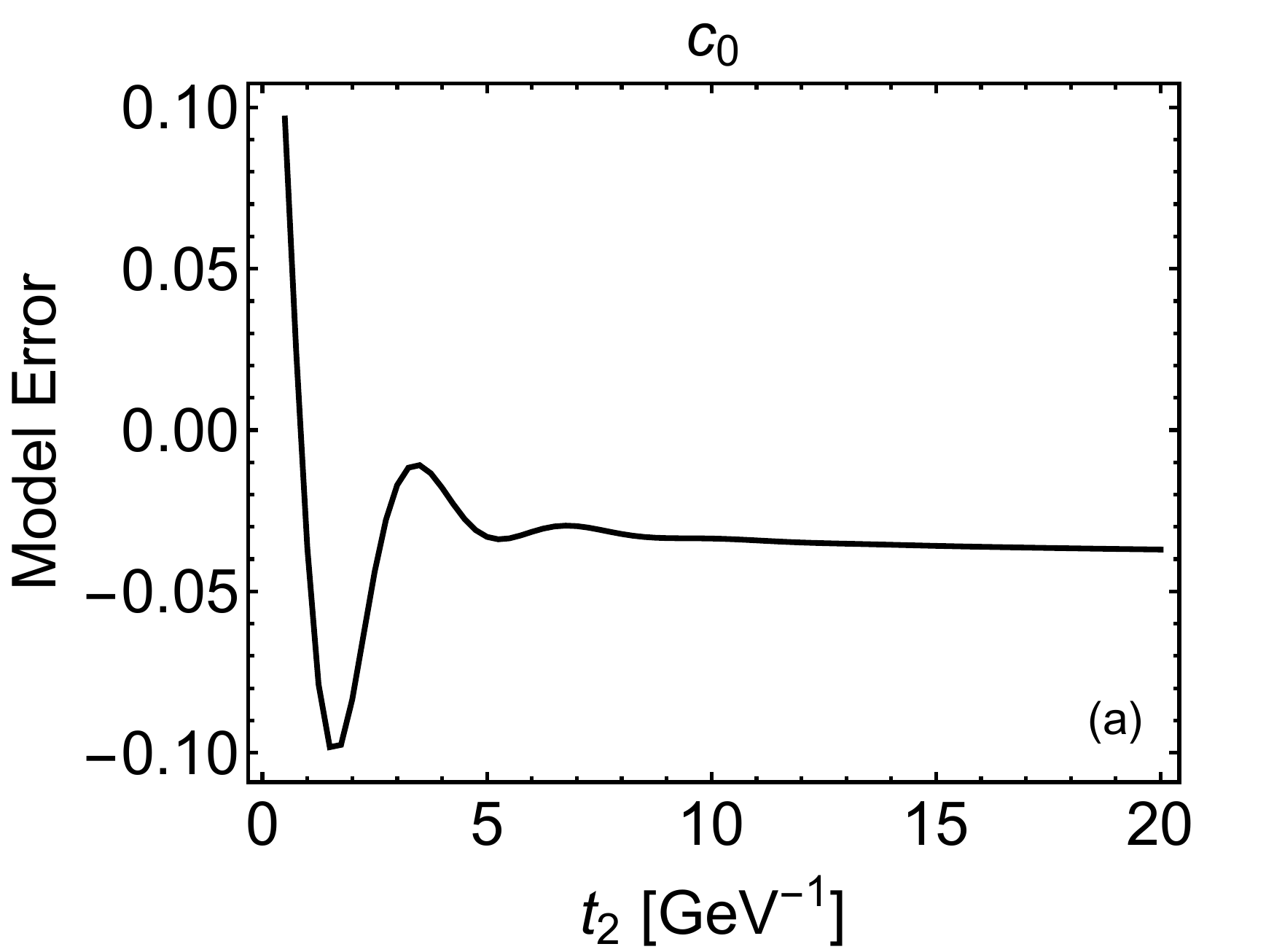}\hspace{5mm}
\includegraphics[width=0.45\linewidth]{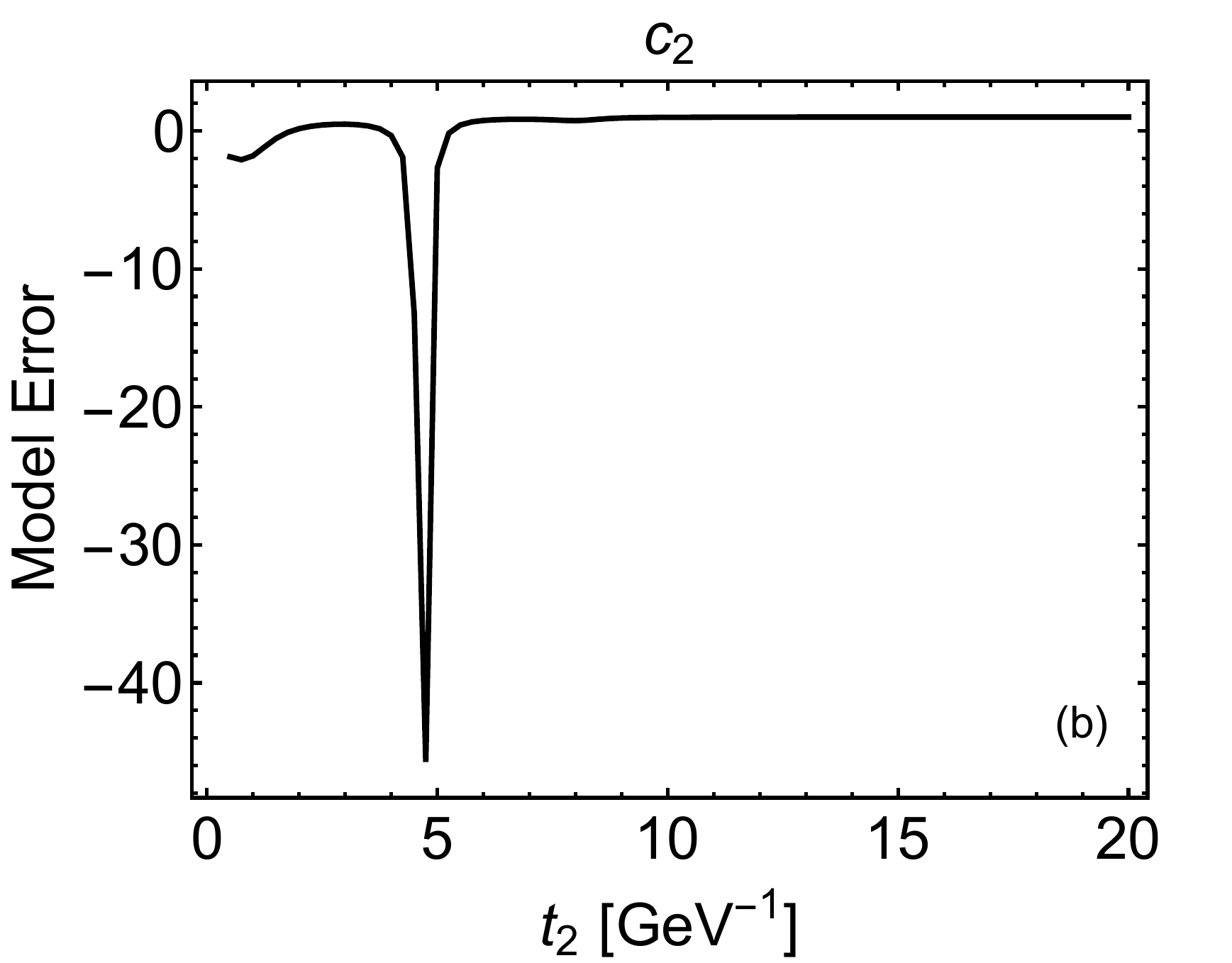}
}
\caption{Model error computed as a function of the transition time $t_2$.  Model error is computed as the difference between the final overlap coefficient squared obtained from the real-time evolution and the adiabatic approximation, divided by the real-time evolution result.  Panel (a) shows the ground state ($n=0$) model error and panel (b) second excited state ($n=2$) model error.}
\label{fig:cserror}
\end{figure*}
 
We begin by making some comparisons using the simpler harmonic oscillator potential.   Our goal is to provide a simple test case where the time-dependent wave functions are known analytically.  For this test case, we took the heavy quark mass to be $m = 1$ GeV and we constructed a piecewise complex-valued spring constant with real-part given by
\be
    \Re(k) = \left\{
        \begin{array}{ll}
            1-\frac{1}{2}\frac{t}{t_{2}} & \quad 0 \leq t \leq t_{2}  \, , \\
            \frac{1}{2} & \quad t > t_{2} \, ,
        \end{array}
    \right. 
\label{eq:cho1}
\ee
and imaginary part given by
\be    
   \Im(k) = \left\{
        \begin{array}{ll}
           \frac{t}{t_{2}}-1 & \quad 0 \leq t \leq t_{2} \, , \\
           0 & \quad t > t_{2} \, .
        \end{array}
    \right. 
    \label{eq:cho2}
\ee
where the implicit units for the real and imaginary parts of the spring constant $k$ are GeV$^3$.  The time scale $t_2$ above can be varied to assess the effect of the speed at which the potential changes on quantum state mixing.  At the same time, in order to mimic the dropping of the temperature in the QGP, we also make the real part change with time on the same time scale, $t_2$.  One expects that the adiabatic approximation should work for large $t_2$ but should fail for small $t_2$.  In the limit $t_2 \rightarrow \infty$ the adiabatic approximation is exact.
 
In Figs.~\ref{fig:cs} and \ref{fig:cserror} we present a comparison of the overlap coefficients obtained using the real-time evolution of the Schr\"odinger equation and the adiabatic approximation~\eqref{eq:adiabaticapprx}.  For the real-time solution we used 1024 points and a one-dimenionsional box size of $L = 40\;{\rm GeV}^{-1}$.  For the temporal step size we use $\Delta t = 5 \times 10^{-4} \; {\rm GeV}^{-1}$.  For the initial condition, we assumed that at $t=0$ that the quantum wave function was a linear combination of the ground state ($n=0$) and the first three excited states ($n=1,2,$ and 3) with equal probability, ie. $c_{0,1,2,3}(0) = 0.5$ with all other overlaps equal to zero.  In the panel (a) of Fig.~\ref{fig:cs} we show the ground state ($n=0$) overlap coefficient squared normalized to its initial value for the case $t_2 = 4.67$ GeV$^{-1}$.  This value of $t_2$ was chosen in order to provide an example case where there are large corrections to the adiabatic approximation.  The solid black line shows the overlap obtained using the real-time method and the dashed red line shows the result obtained in the adiabatic approximation.   From this figure we see that, for this particular choice of $t_2$, the ground state survival probability is very well described by the adiabatic approximation, with the deviation between the two results being approximately 3\%.  In contrast, in Fig.~\ref{fig:cserror}(b) we plot the same quantities for the second excited state ($n=2$).  As this figure demonstrates, in the case of the excited state there exist values of $t_2$ for which the adiabatic approximation can fail dramatically.

To investigate the error made by the adiabatic approximation further, in Fig.~\ref{fig:cserror} we plot the model error computed as the difference between the real-time evolution method and adiabatic approximation divided by the result obtained from the real-time evolution method.  Panel (a) shows the ground state ($n=0$) model error and panel (b) second excited state ($n=2$) model error.  From Fig.~\ref{fig:cserror}(a) we see that for small $t_2$ there can be up to 10\% corrections and, as $t_2$ is increased we see a reduction in the error.  In contrast, In Fig.~\ref{fig:cserror}(b) we see that for a particular choice of $t_2$, namely the value of $t_2 = 4.67$ GeV$^{-1}$ which was used for the examples shown in Figs.~\ref{fig:cs}, we see a huge correction.  This is due to the natural oscillations of different quantum modes.  For this particular choice of $t_2$ the imaginary part of the potential goes to zero during one of these oscillations.  We do not want to overemphasize this case, however, because it's not clear how to draw quantitative conclusions from this toy model for the application to heavy quark suppression.  For this reason, in the next section we will present our results using a realistic complex heavy-quark potential.

\subsection{CKMS potential results}

We made several case studies using the CKMS potential \eqref{eq:ckmsdef}.  In all cases, we assume that the system is undergoing boost-invariant and transversally homogenous one-dimensional Bjorken flow \cite{Bjorken:1982qr} such that
\be
T(\tau) = T_0 \left( \frac{\tau_0}{\tau} \right)^{1/3} \, ,
\label{eq:bjorken}
\ee
where $T_0$ is the initial temperature at the proper time $\tau_0$.  We will keep the initial time fixed to be $\tau_0 = 1 \; {\rm GeV}^{-1} = 0.197$ fm/c and vary the initial temperature to mimic what is seen in a typical Glauber temperature profile in a LHC 2.76 TeV Pb-Pb collision.  In all cases shown below, we terminated the evolution when the temperature fell below the freeze-out temperature $T_{\rm fo} = $150 MeV.  For the CKMS potential, we used a box size of $L = 40 \; {\rm GeV}^{-1}$ and $N=1024$ grid points.  The spatial step size was $\Delta x = L/(N+1)$ with $r=0$ excluded.  The temporal step size was $\Delta \tau = 1 \times 10^{-3} \; {\rm GeV}^{-1}$.

\subsubsection{Bottomonia}

\begin{figure*}[t!]
\centerline{
\includegraphics[width=0.45\linewidth]{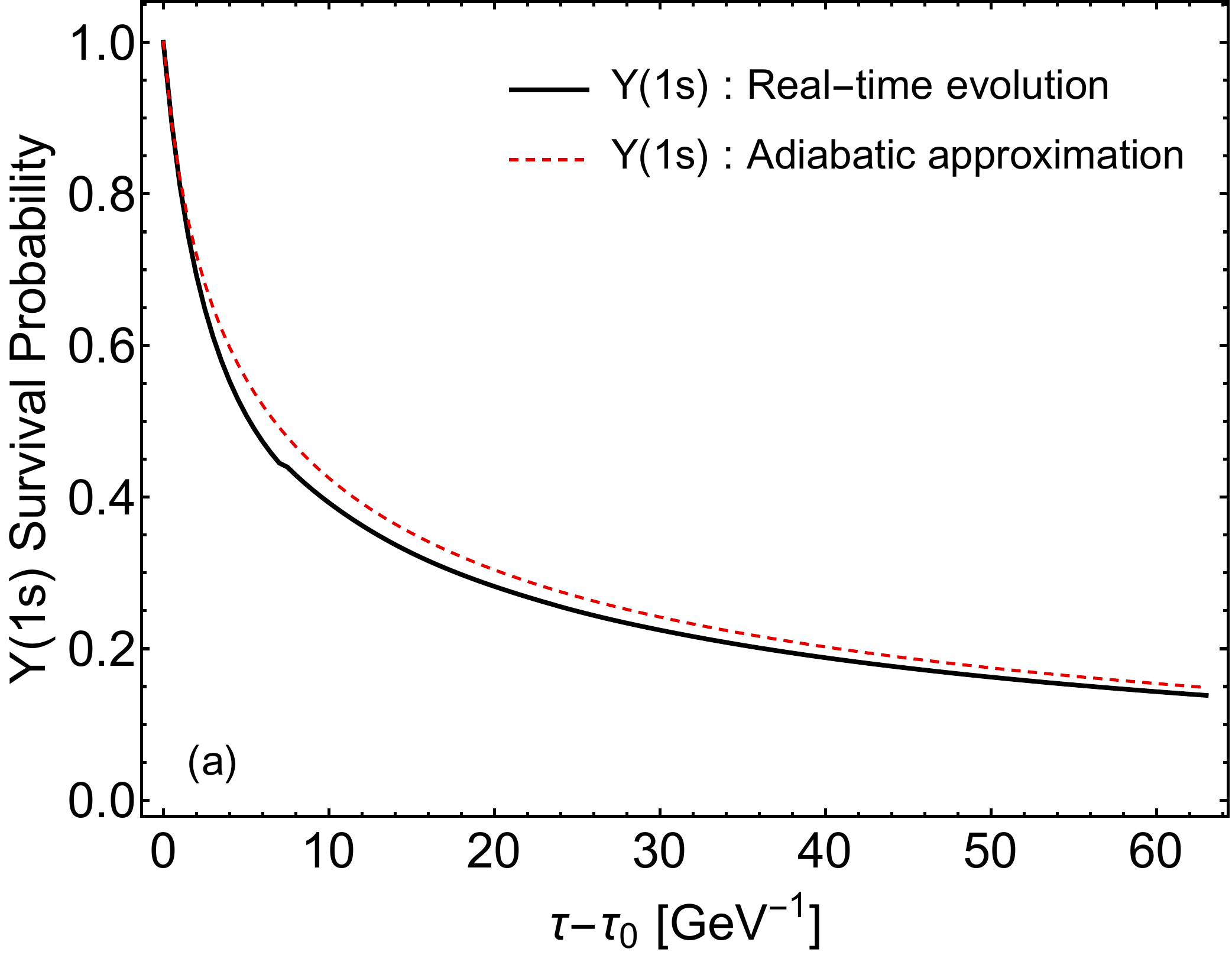}\hspace{5mm}
\includegraphics[width=0.475\linewidth]{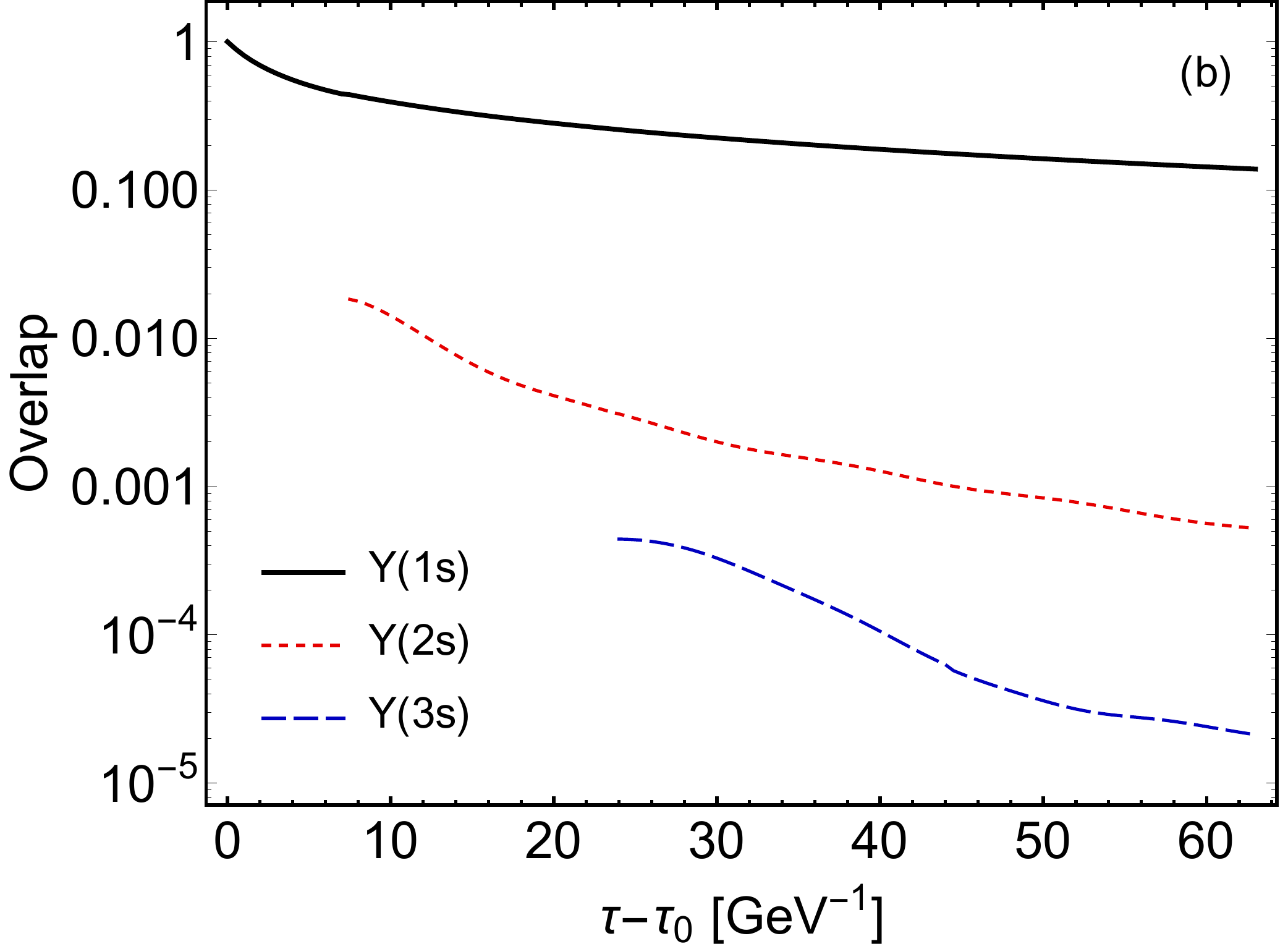}
}
\caption{Panel (a) shows the survival probability of the $\Upsilon(1s)$ as a function of proper time.  In this panel, the solid black line is the result obtained using real-time evolution and the dashed red line is the result obtained using the adiabatic approximation.  Panel (b) shows the overlap coefficients squared extracted using the real-time evolution.  In this panel, the solid black, shorted-dashed red, and long-dashed blue lines are the overlaps computed for the 1s, 2s, and 3s states.  In both panels the initial temperature was taken to be $T_0 = 0.6$ GeV at  $\tau_0 = $ 1 GeV$^{-1}$ and the initial wave function consisted of a 1s state with the wave function determined self-consistently using the CKMS potential.}
\label{fig:Y1s-comp-0p6}
\end{figure*}
 
\begin{figure*}[t!]
\centerline{
\includegraphics[width=0.45\linewidth]{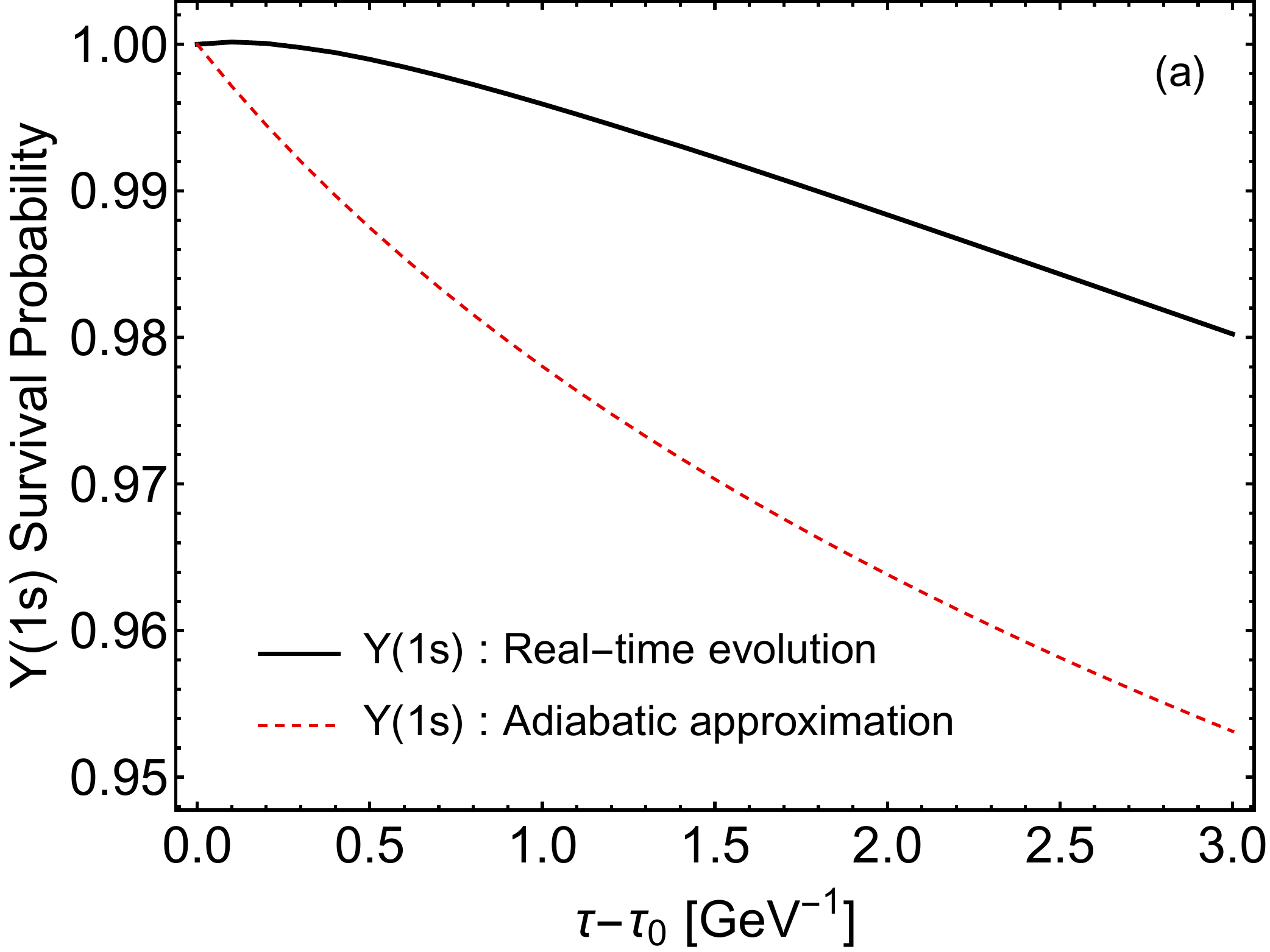}\hspace{5mm}
\includegraphics[width=0.45\linewidth]{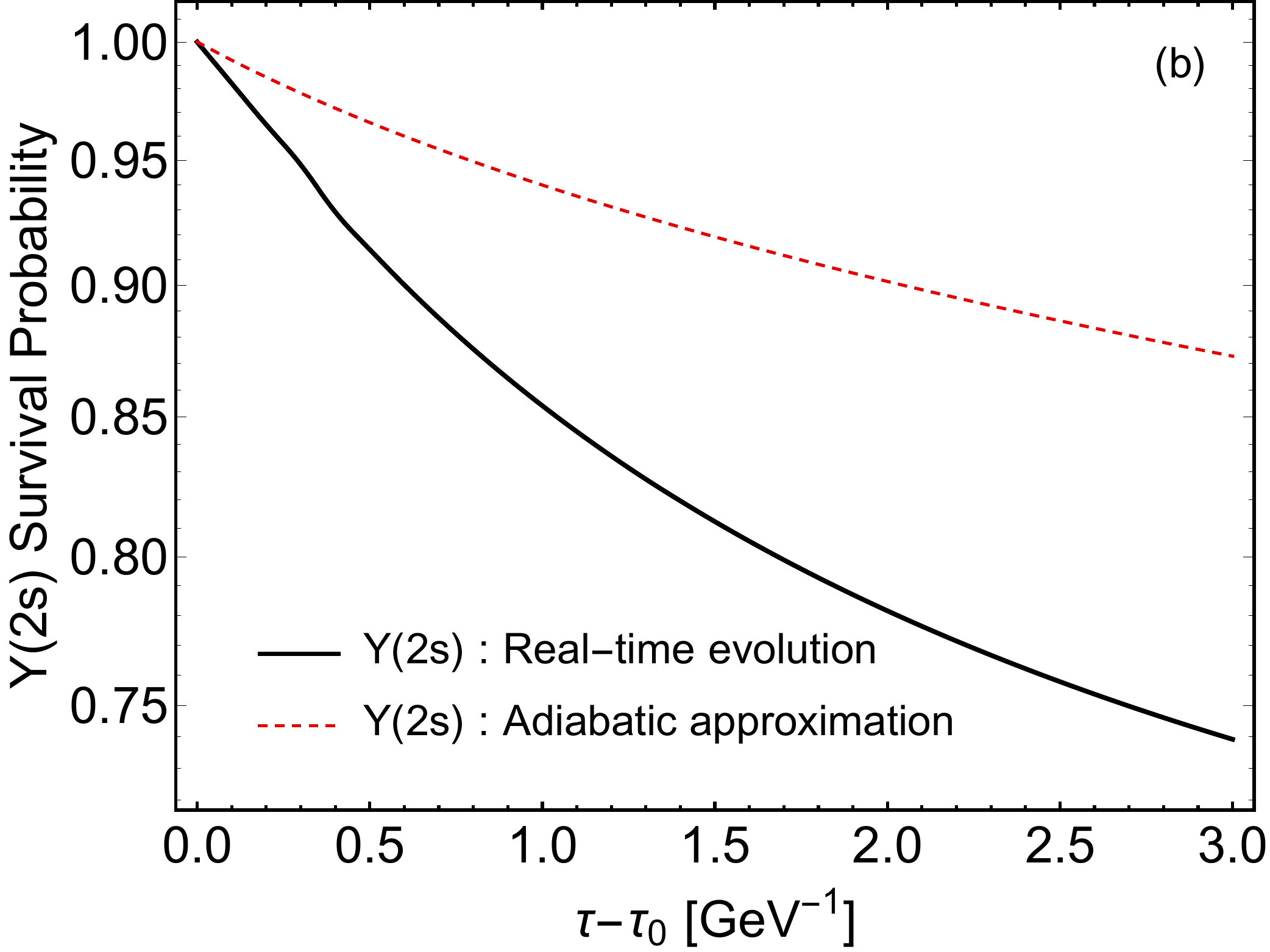}
}
\vspace{5mm}
\centerline{
\includegraphics[width=0.465\linewidth]{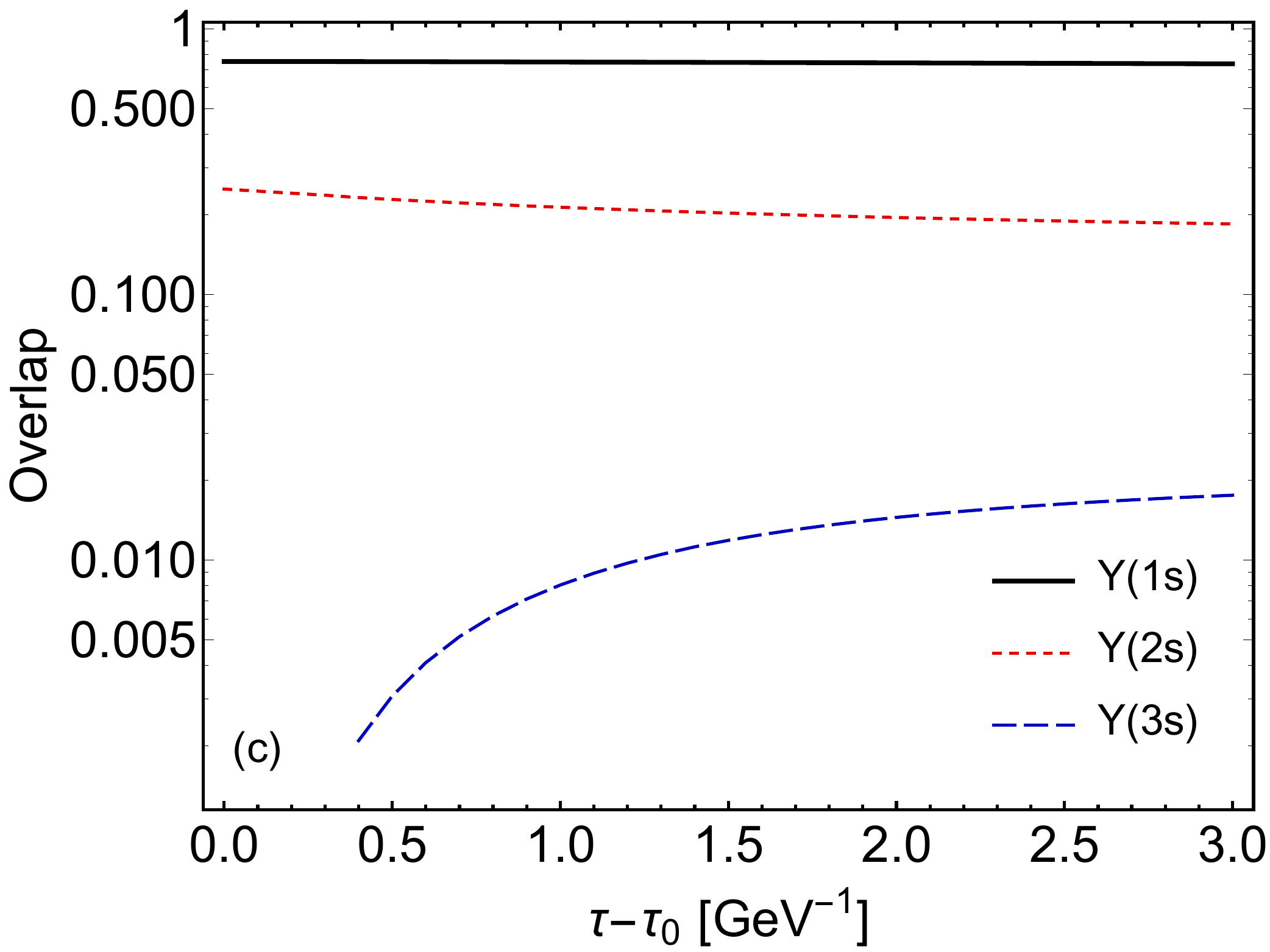}
}
\caption{Panels (a) and (b) show the survival probability of the $\Upsilon(1s)$ and $\Upsilon(2s)$ as a function of proper time.  In these panels, the solid black line is the result obtained using real-time evolution and the dashed red line is the result obtained using the adiabatic approximation.  Panel (c) shows the overlap coefficients squared extracted using the real-time evolution.  In this panel, the solid black, shorted-dashed red, and long-dashed blue lines are the overlaps computed for the 1s, 2s, and 3s states.  In both panels the initial temperature was taken to be $T_0 = 0.225$ GeV at  $\tau_0 = $ 1 GeV$^{-1}$.  The initial wave function consisted of a linear combination of 1s and 2s components with the wave functions determined self-consistently using the CKMS potential.}
\label{fig:Y1s-comp-0p225}
\end{figure*}
 
\begin{figure*}[t!]
\centerline{
\includegraphics[width=0.45\linewidth]{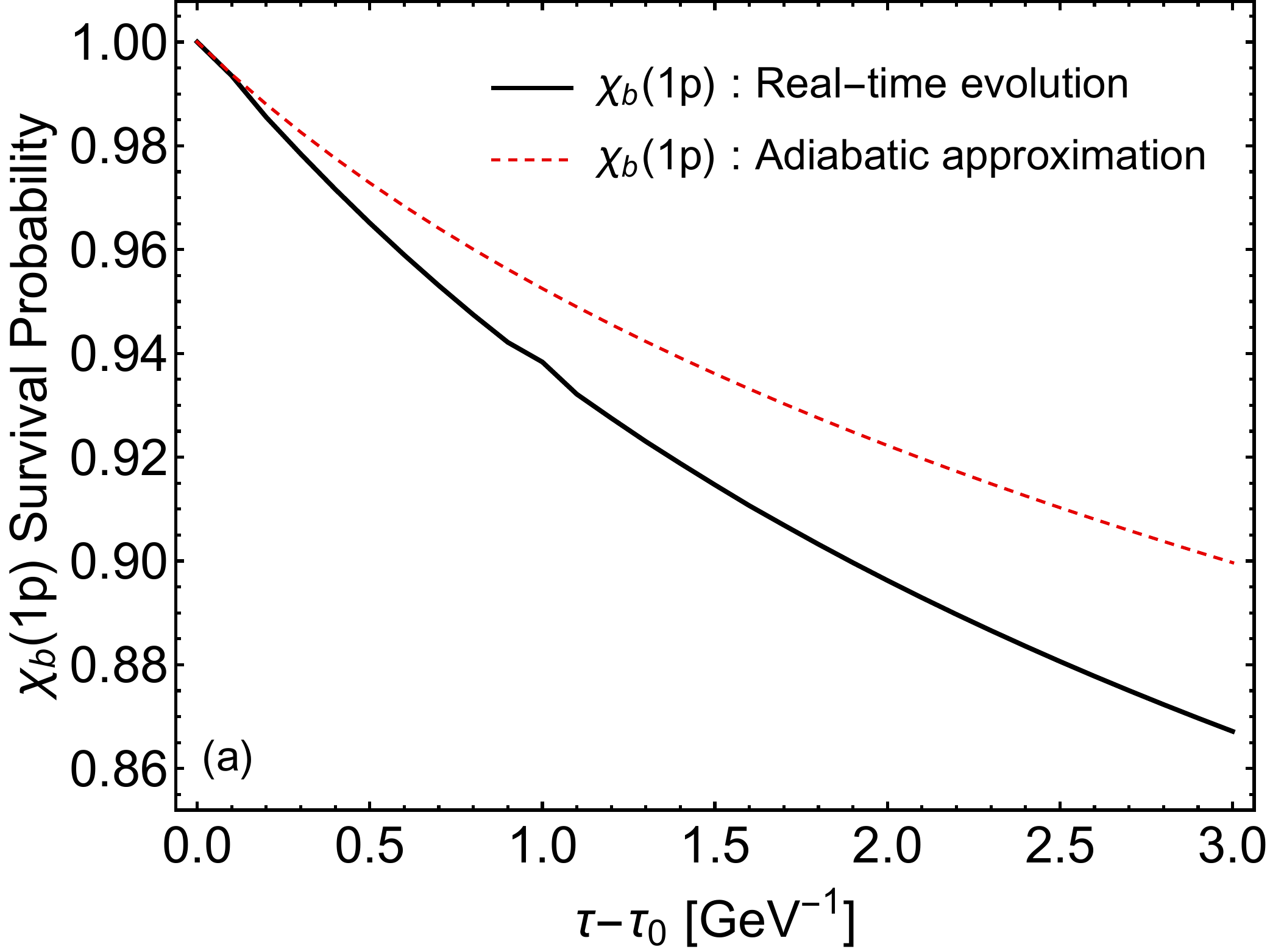}\hspace{5mm}
\includegraphics[width=0.465\linewidth]{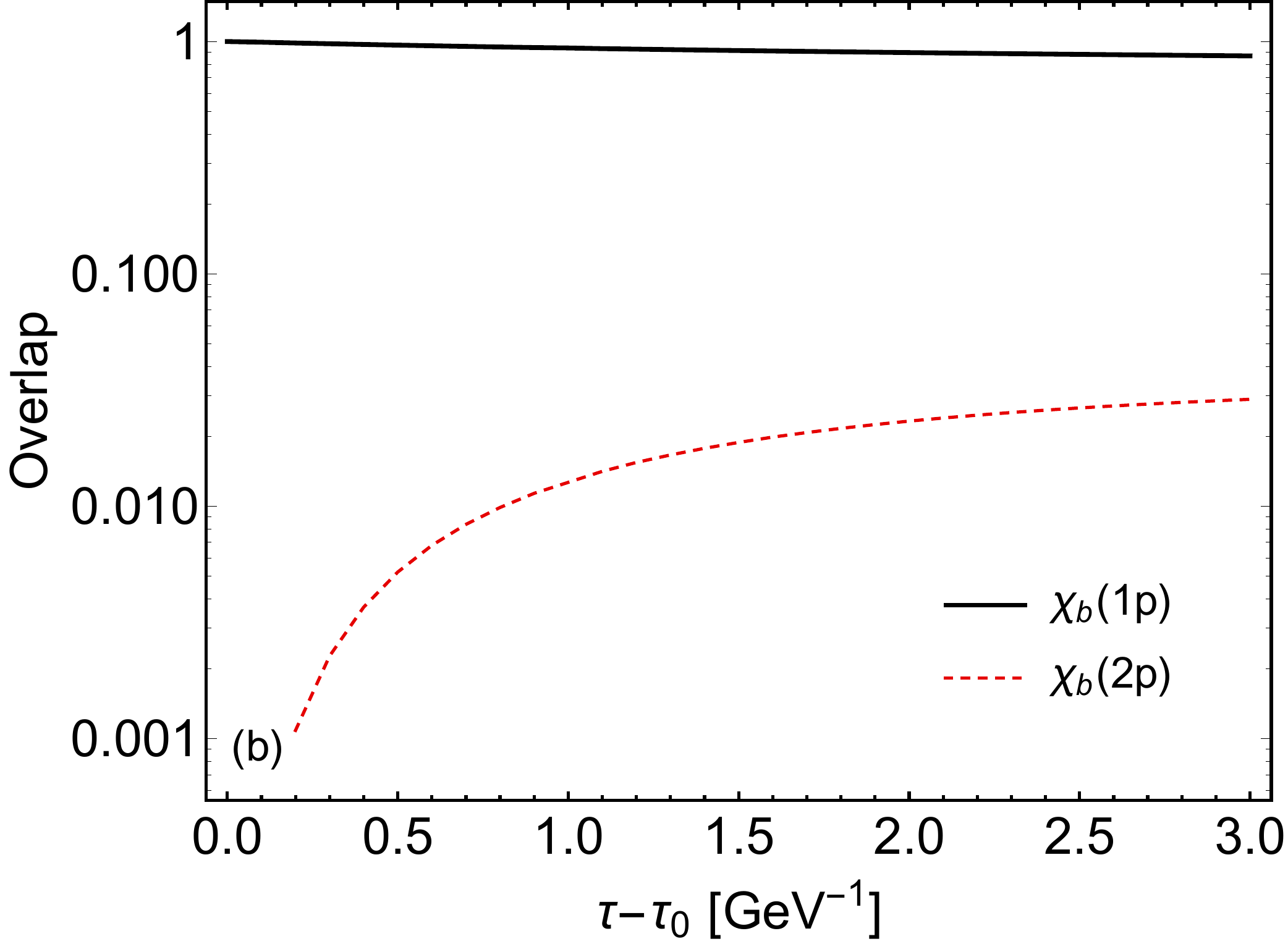}
}
\caption{Panel (a) shows the survival probability of the $\chi_b(1p)$ as a function of proper time.  Panel (b) shows the overlap coefficients squared extracted using the real-time evolution.  In this panel, the solid black and shorted-dashed red lines are the overlaps computed for the 1p and 2p states, respectively.  In both panels the initial temperature was taken to be $T_0 = 0.6$ GeV at  $\tau_0 = $ 1 GeV$^{-1}$ and the initial wave function consisted of a 1s state with the wave function determined self-consistently using the CKMS potential.}
\label{fig:Xb1p-comp-0p225}
\end{figure*}

In Figs.~\ref{fig:Y1s-comp-0p6}-\ref{fig:Xb1p-comp-0p225} we collect our results for bottomonium states.  In Fig.~\ref{fig:Y1s-comp-0p6} we plot the survival probability of the $\Upsilon(1s)$ state as a function of proper time in panel (a) and the extracted probabilities $|c_n(t)|^2$ for the $1s$, $2s$, and $3s$ states in panel (b).  For Fig.~\ref{fig:Y1s-comp-0p6} we assumed an initial temperature of $T_0 = 0.6$ GeV, which is the temperature generated in the center of the QGP in a central Pb-Pb collision at LHC energies \cite{Alqahtani:2017jwl}. For the initial wave function, we took a pure $\Upsilon(1s)$ wave function which was determined self-consistently at the initial temperature of the QGP using the CKMS potential.  Finally, for the bottom quark mass we used $m_b = 4.7$ GeV which is the pole mass reported in the Particle Data Group (PDG) listings \cite{PhysRevD.98.030001}.

In panel (a) of Fig.~\ref{fig:Y1s-comp-0p6} the black solid line is the result of extracting the $\Upsilon(1s)$ overlap probability from the real-time solution and the dashed line is the result obtained using the adiabatic approximation \eqref{eq:adiabaticapprx}.  This panel demonstrates that there are only small corrections to the $\Upsilon(1s)$ survival probability coming from quantum state mixing in the center of the QGP.  At the final time shown, the two results differ by approximately 6\%.  We do see, however, that the real-time evolution predicts enhanced 1s suppression compared to the adiabatic approximation.  

Turning to panel (b) of Fig.~\ref{fig:Y1s-comp-0p6} we see quantum state mixing due to the time evolution of the potential which results in the formation of $\Upsilon(2s)$ and $\Upsilon(3s)$ states as the QGP cools.  In this panel, the initial times for the 2s and 3s curves are determined by the point in proper time at which these states become bound in the QGP, as determined by the point-and-shoot method described in Sec.~\ref{sec:cpas}.  Despite the presence of quantum state mixing, in this case, however, we see only a maximum overlap of approximately 0.02 with the 2s state and 0.0004 with the 3s state.

In Fig.~\ref{fig:Y1s-comp-0p225} we plot the same quantities as in Fig.~\ref{fig:Y1s-comp-0p6}, however, in this case we assumed an initial temperature of $T_0 = 0.225$ GeV.  At this initial temperature both the 1s and 2s states are bound, so we take the initial wave function to be a linear combination of the 1s and 2s states with a relative probability of 33\% for the 2s.  This relative probability is set according to the ratio of the 2s to 1s total production cross sections in pp collisions at 2.76 TeV collision energy, which are 48 nb and 145 nb for $|y| < 2.4$, respectively \cite{Khachatryan:2016xxp}.  As can be seen from panel (a) of Fig.~\ref{fig:Y1s-comp-0p225}, for this initial condition there is clearly a difference in the evolution and final results for the $\Upsilon(1s)$ suppression, however, numerically this represents only a 3\% difference between the two results shown at the final time.  One other thing we notice from panel (a) is that the ordering of the curves is reversed compared to panel (a) of Fig.~\ref{fig:Y1s-comp-0p6}, with the real-time solution resulting in less suppression at this lower temperature.   

In panel (b) of Fig.~\ref{fig:Y1s-comp-0p225}, we present a comparison of the real-time evolution result for the $\Upsilon(2s)$ survival probability with the adiabatic approximation.  As can be seen from this panel, there are larger corrections to the adiabatic approximation when considering the $\Upsilon(2s)$, with the final difference between the two approaches being approximately 18\% with the real-time evolution method predicting larger suppression.  In panel (c) of Fig.~\ref{fig:Y1s-comp-0p225}, we present the combined evolution of the overlaps with the 1s, 2s, and 3s states as a function of proper time.  Once again we see formation of excited states that were not in the initial wave function.  In this case, in contrast to Fig.~\ref{fig:Y1s-comp-0p6}(b) we see that the 3s overlap increases monotonically with time.  Even with this, one still finds a rather small 3s overlap which is approximately 0.02 at the final time.

To close out this subsection, we turn to Fig.~\ref{fig:Xb1p-comp-0p225} which shows results for the $\chi_b$(1p) and $\chi_b$(2p) states.  The panels are the same as the previous figures and, in this case, for the initial condition we used a pure $\chi_b(1p)$ state determined self-consistently at the assumed initial temperature of $T_0 = 0.225$ GeV.   From panel (a) we see that there is an approximately 4\% difference between the final $\chi_b$(1p) suppression when comparing the real-time evolution and the adiabatic approximation.  In panel (b) we once again see an excited state $\chi_b(1p)$ overlap which increases monotonically in time, but it remains small, with the maximum overlap being approximately 0.03 at the final time.

\subsubsection{Charmonia}

\begin{figure*}[t!]
\centerline{
\includegraphics[width=0.45\linewidth]{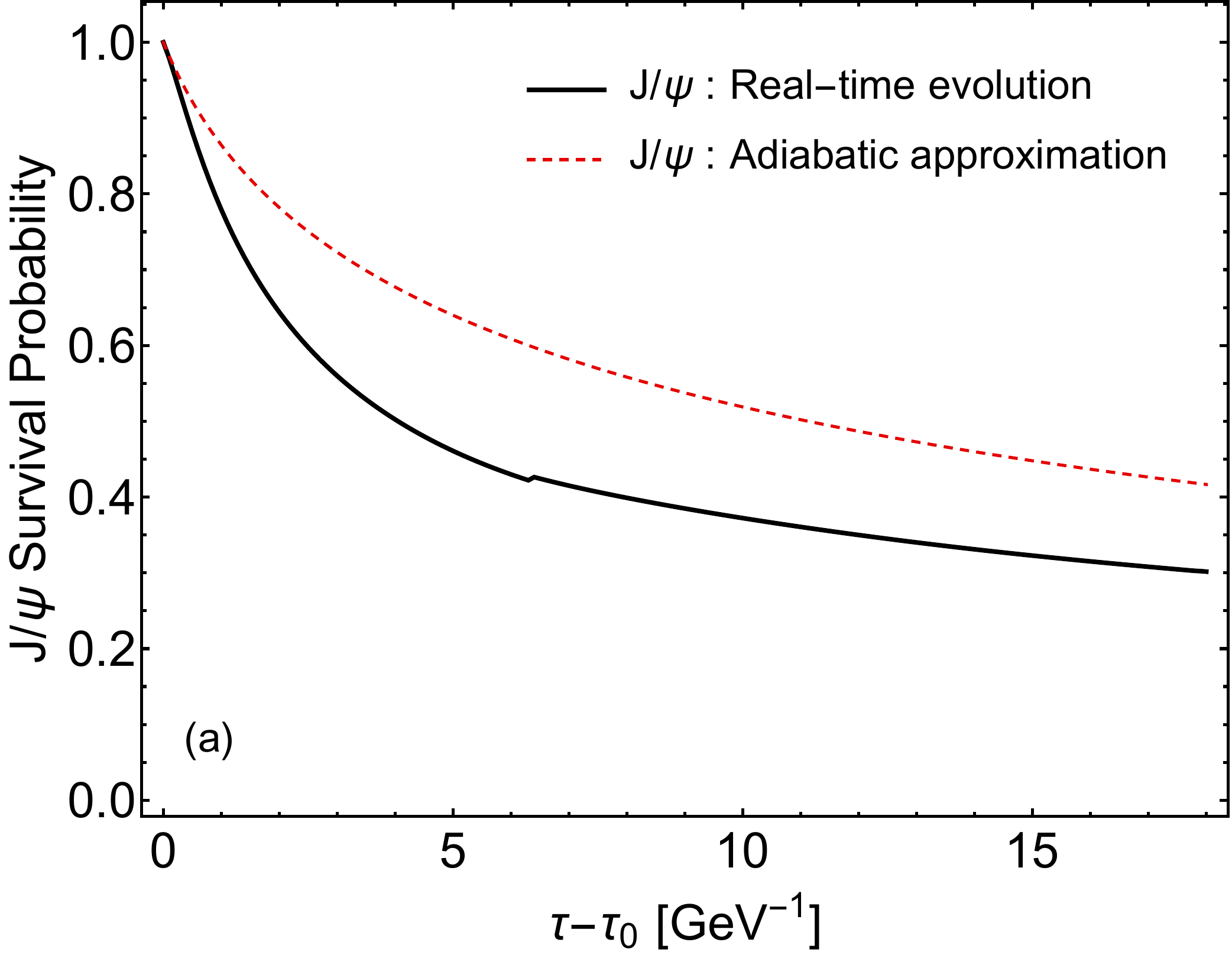}\hspace{5mm}
\includegraphics[width=0.475\linewidth]{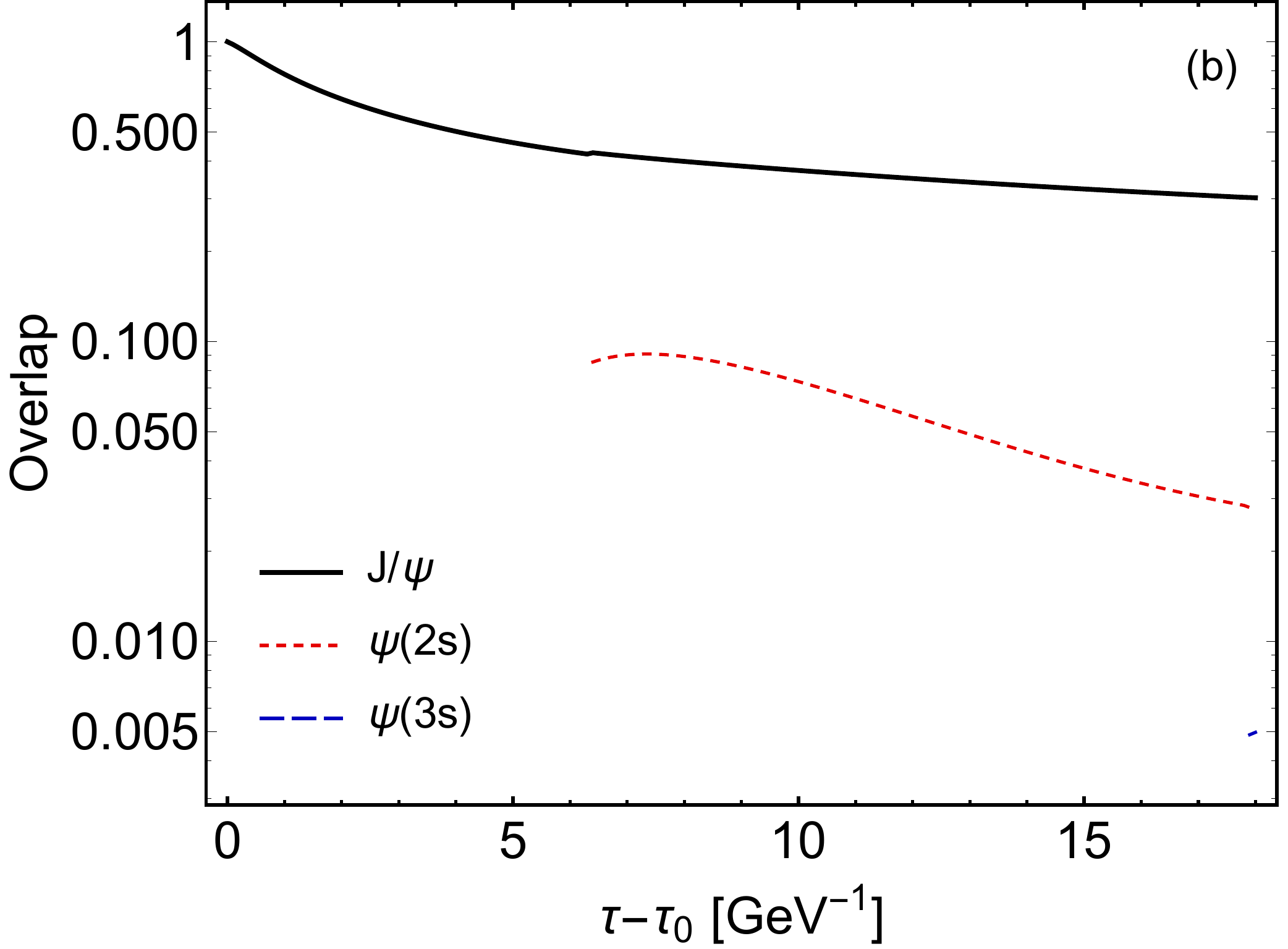}
}
\caption{Panel (a) shows the survival probability of the $J/\psi$ as a function of proper time.  In this panel, the solid black line is the result obtained using real-time evolution and the dashed red line is the result obtained using the adiabatic approximation.  Panel (b) shows the overlap coefficients squared extracted using the real-time evolution.  In this panel, the solid black, shorted-dashed red, and long-dashed blue lines are the overlaps computed for the 1s, 2s, and 3s states.  In both panels the initial temperature was taken to be $T_0 = 0.4$ GeV at  $\tau_0 = $ 1 GeV$^{-1}$ and the initial wave function consisted of a 1s state with the wave function determined self-consistently using the CKMS potential.}
\label{fig:JPsi-comp-0p4}
\end{figure*}

In this subsection, we consider charmonia.  The only change made to the model parameters was to change the heavy quark mass to the charm pole mass of $m_c = 1.7$ GeV taken from the PDG \cite{PhysRevD.98.030001}.  In Fig.~\ref{fig:JPsi-comp-0p4} we show the comparison between the adiabatic approximation and the real-time evolution in panel (a) and a plot of the extracted overlaps in panel (b).  The initial temperature was taken to be $T_0 = 0.4$ GeV and the initial wave function was self-consistently obtained using the point-and-shoot method with the CKMS potential.  This temperature corresponds to the central temperature achieved in $b =$ 12 fm Pb-Pb collisions at 2.76 TeV.  At this initial temperature, the only surviving $s$-wave state is the $1s$ $J/\psi$ state.  

As panel (a) of  Fig.~\ref{fig:JPsi-comp-0p4} demonstrates, we observe a larger difference between the real-time evolution result and the adiabatic approximation for the $J/\Psi$ using this initial temperature, with the difference between the two results being approximately 36\% at the final time.  In panel (b) we show the numerically extracted overlaps with the $J/\Psi$, $\Psi(2s)$, and $\Psi(3s)$ as a function of proper time.  As can be seen from this figure, the 2s and 3s overlaps are approximately 0.03 and 0.005, respectively.

\begin{figure*}[t!]
\centerline{
\includegraphics[width=0.45\linewidth]{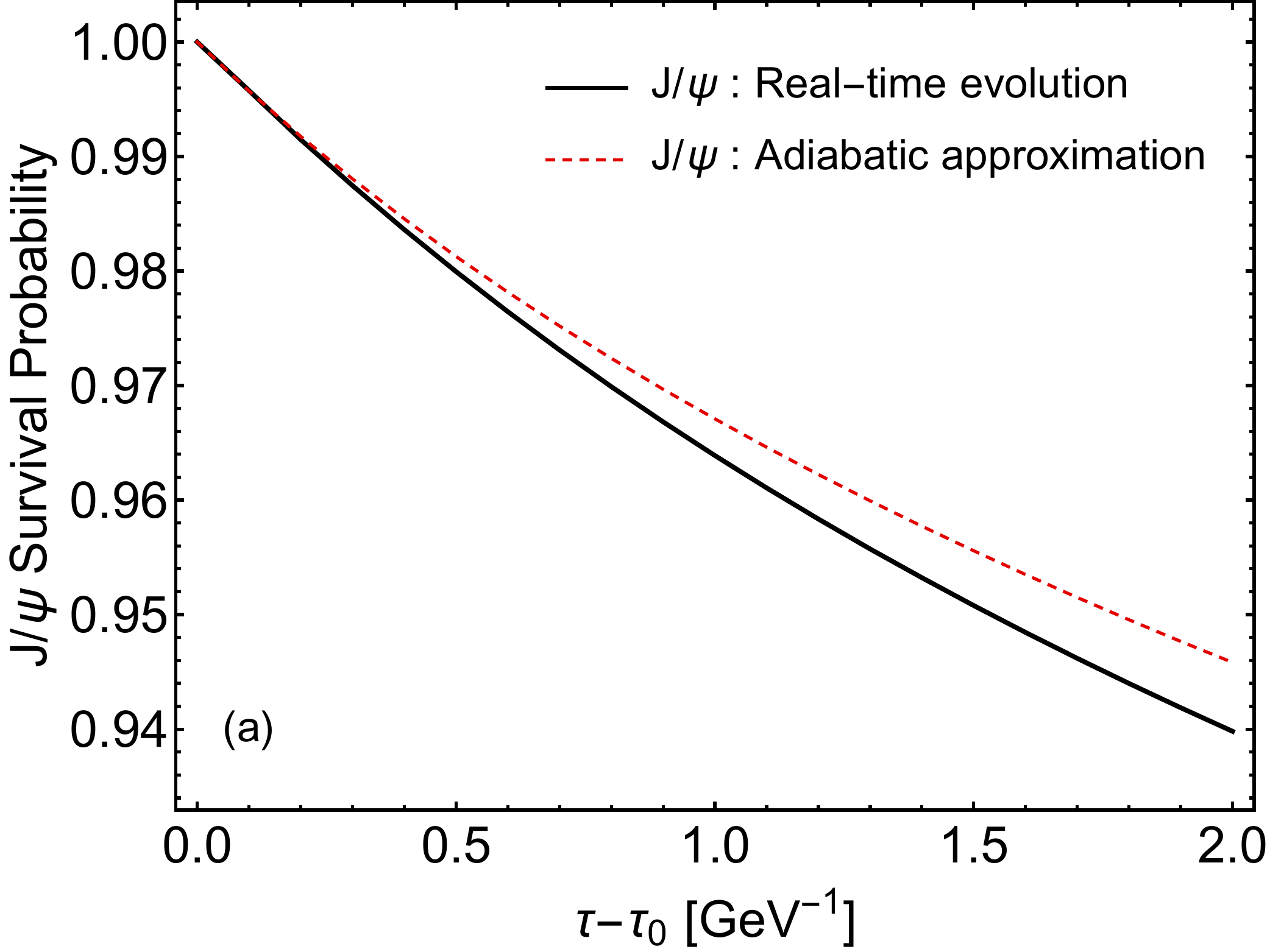}\hspace{5mm}
\includegraphics[width=0.46\linewidth]{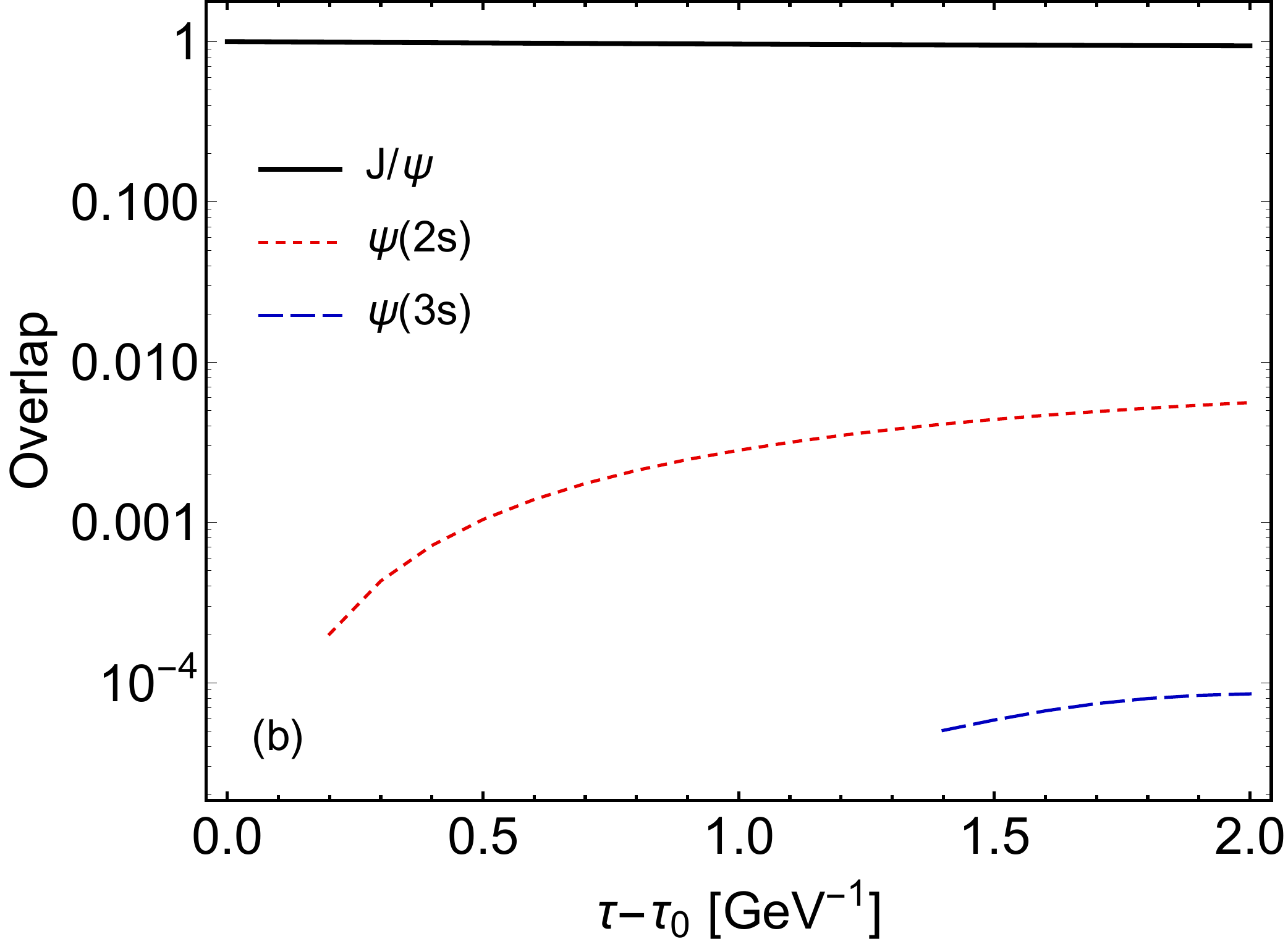}
}
\caption{Same as Fig.~\ref{fig:JPsi-comp-0p4}, but with $T_0 = 0.2$ GeV.}
\label{fig:JPsi-comp-0p2}
\end{figure*}

Next, we consider Fig.~\ref{fig:JPsi-comp-0p2} in which we have used a lower initial temperature of $T_0 = $ 0.2 GeV.  At this lower initial temperature the only $s$-wave bound state is still the $J/\Psi$ so we initialized the wave function as a pure $J/\Psi$ state using the in-medium point-and-shoot method.  For this initial temperature we see from panel (a) of Fig.~\ref{fig:JPsi-comp-0p2} that the difference between the real-time and adiabatic results is approximately 0.6\%.  The overlaps with the 1s, 2s, and 3s states are shown in panel (b) of Fig.~\ref{fig:JPsi-comp-0p2}.  From this figure we observe that both the 2s and 3s overlaps are monotonically increasing with the overlap being approximately 0.006 and 0.0001, respectively, at the final time.

\begin{figure*}[t!]
\centerline{
\includegraphics[width=0.45\linewidth]{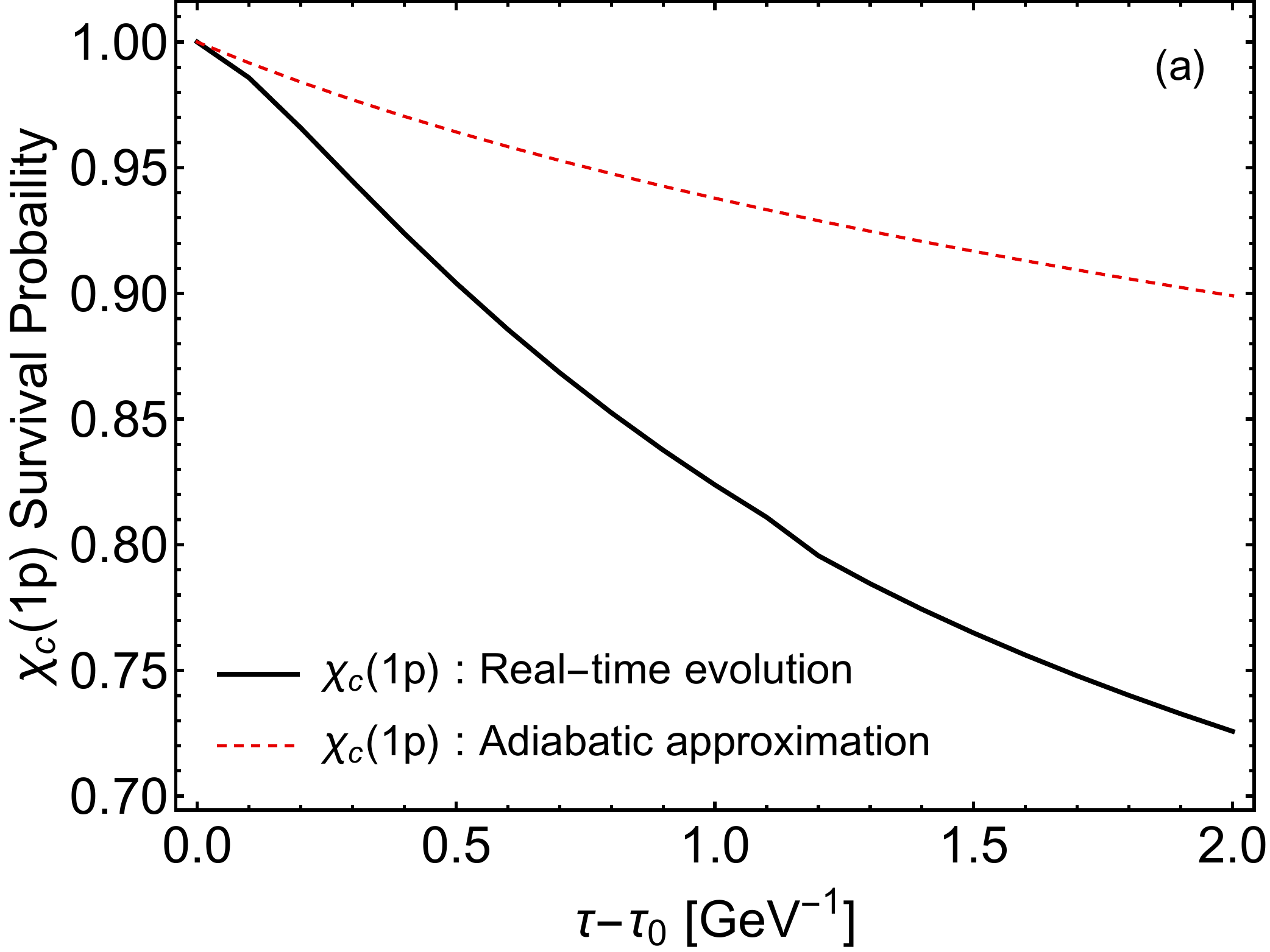}\hspace{5mm}
\includegraphics[width=0.45\linewidth]{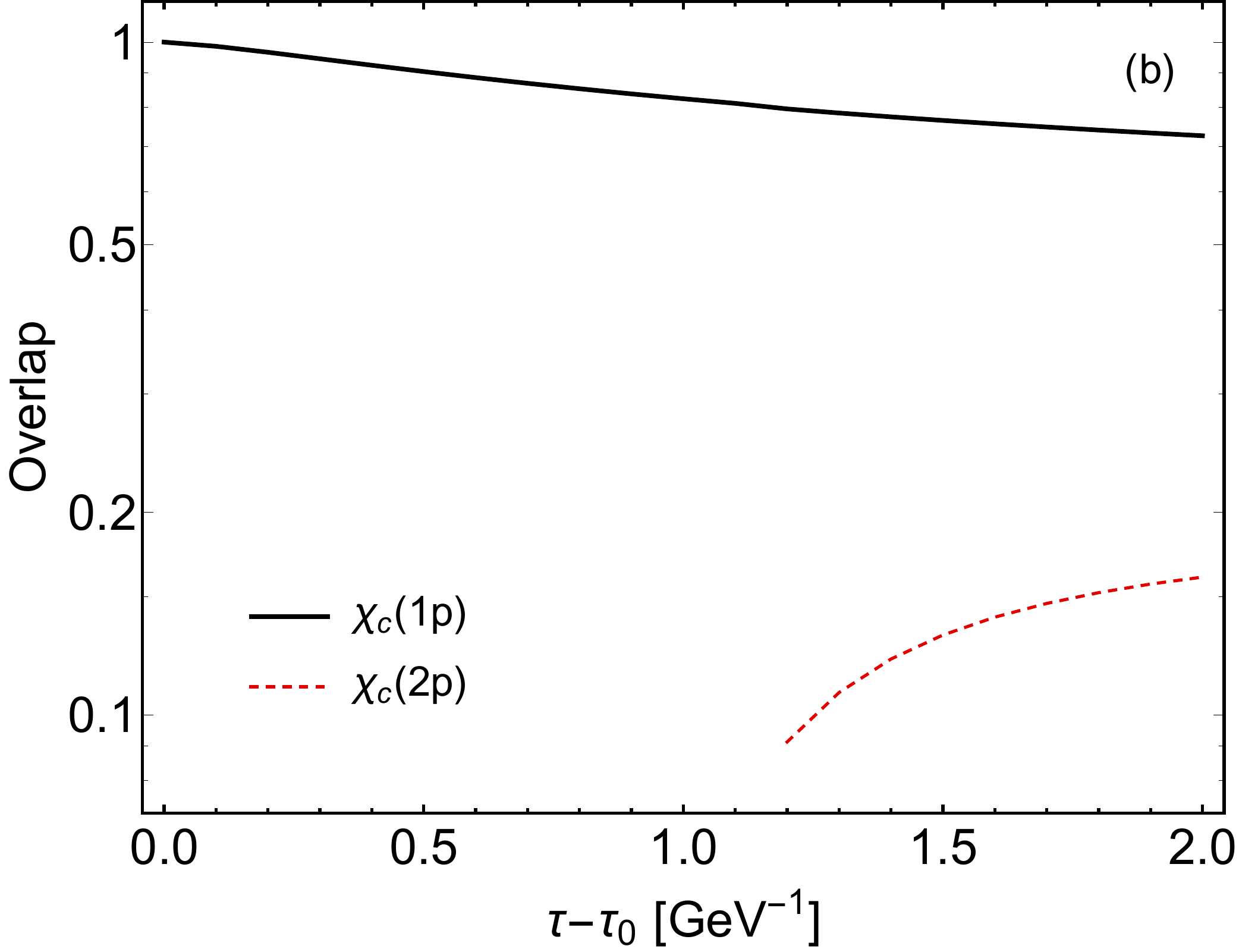}
}
\caption{Panel (a) shows the survival probability of the $\chi_c(1p)$ as a function of proper time.  In this panel, the solid black line is the result obtained using real-time evolution and the dashed red line is the result obtained using the adiabatic approximation.  Panel (b) shows the overlap coefficients squared extracted using the real-time evolution.  In this panel, the solid black and shorted-dashed red lines are the overlaps computed for the 1p and 2p states.  In both panels the initial temperature was taken to be $T_0 = 0.2$ GeV at  $\tau_0 = $ 1 GeV$^{-1}$ and the initial wave function consisted of a 1p state with the wave function determined self-consistently using the CKMS potential.}
\label{fig:Xc1p-comp-0p2}
\end{figure*}

Finally, in Fig.~\ref{fig:Xc1p-comp-0p2} we present the survival probability and overlaps for $p$-wave charmonium states in panels (a) and (b), respectively.  The initial temperature was the same as in Fig.~\ref{fig:JPsi-comp-0p2}, namely $T_0 = 0.2$ GeV.   As panel (a) demonstrates, in this case one sees a larger deviation between the real-time and adiabatic approximation, with the two results differing by approximately 23\%.  This difference is similar to that seen in Fig.~\ref{fig:Y1s-comp-0p225}(b) for the $\Upsilon(2s)$.  Turning to panel (b) of  Fig.~\ref{fig:Xc1p-comp-0p2}, we see that the overlap with the 2p state is monotonically increasing with time and is approximately 0.16 at the final time.

To close out this section in Table \ref{tab:sum} we summarize the results of the effect on the survival probability at the final time.  The left column shows the initial temperature used in GeV, the middle column shows the state considered, and the third column shows the final effect on the primordial survival probability, $R_{AA}$ (ignoring feed down effects).  Negative numbers in the third column indicate that $R_{AA}$ is decreased (more suppresion) while positive numbers indicate the opposite.

\begin{table}
\small
\begin{center}
{
\def\arraystretch{1.3}
\setlength{\tabcolsep}{0.4em}
\begin{tabular}{|c|c|c|}
    \hline 
    \bf $\bf T_0$ [GeV] & \bf State & \bf \% correction to primordial $\bf R_{AA}$\\
    \hline \hline
    0.6 & $\Upsilon(1s)$ & -6\% \\
    \hline \hline
    \multirow{3}{*}{0.225} & $\Upsilon(1s)$ & +3\% \\ \cline{2-3}
    & $\Upsilon(2s)$ & -18\% \\  \cline{2-3}
    & $\chi_b(1p)$ & -4\% \\
    \hline \hline
    0.4 & $J/\Psi$ & -36\% \\ 
    \hline \hline
    \multirow{2}{*}{0.2} & $J/\Psi$ & -0.6\% \\ \cline{2-3}
    & $\chi_c(1p)$ & -23\% \\  \hline
\end{tabular}
}
\end{center}
\caption{Summary of the CKMS potential test cases considered herein.  The quoted percentage change is computed at the final time of evolution and does not take into account feed down corrections.  Positive corrections correspond to reduced suppression and negative corrections to enhanced suppression.}
\label{tab:sum}
\end{table}

\section{Discussion and Conclusions}
\label{sec:conclusions}

In this paper, we presented the results of the real-time evolution of heavy quarkonium state subject to a time-dependent complex-valued potential.  We considered two cases:  a complex harmonic oscillator (CHO) potential and a complex Karsch-Mehr-Satz (CKMS) potential.  The first potential was used as a simple test cases since, in this case, one can determine the eigenfunctions and corresponding eigenvalues analytically by a simple analytic continuation of the solutions obtained for real-valued spring constants.  We presented results for the survival probability extracted by taking projections of the in-medium bound state wave functions with the wave function obtained from numerical solution of the time-dependent Schr\"odinger equation using the ``split-operator'' or ``pseudospectral'' method.  To assess the impact of the speed of change of the potential we introduced a complex spring constant whose real value interpolated between fixed values \eqref{eq:cho1} and whose imaginary part interpolated between a finite value and zero \eqref{eq:cho2} on a variable time scale $t_2$.  As expected, we found that, as the speed of the transition was reduced, the adiabatic approximation became more reliable, however, we found that there were particular choices of $t_2$ which could cause large deviations from the adiabatic approximation.  

After this basic example was presented, we then moved on to assess the impact of going beyond the adiabatic approximation using a realistic heavy quarkonium potential \eqref{eq:ckmsdef}.  For this purpose, we assumed that the temperature evolved in time using simple boost-invariant and transversally homogeneous Bjorken evolution \eqref{eq:bjorken}.  For the bottomonium system we considered initial temperatures of $T_0 = 0.6$ GeV and $T_0 = 0.225$ GeV, with former being the typical temperature achieved in a central Pb-Pb collision at 2.76 TeV/nucleon collision energy and the latter being either the temperature achieved in the periphery of a central Pb-Pb collision or in the center of a peripheral Pb-Pb collision with an impact parameter of $b = $ 14.7 fm.  For the higher temperature, we found that the corrections to the adiabatic approximation for the $\Upsilon(1s)$ survival probability were on the few percent level, however, for the lower temperature, we found that, although the $\Upsilon(1s)$ still did not receive large corrections, the $\Upsilon(2s)$ exhibited a larger correction when the full real-time evolution was employed.

For charmonium, we analyzed two initial temperatures, $T_0 = 0.4$ GeV and $T_0 = 0.2$ GeV, which were chosen such that, at the higher initial temperature, only the $J/\Psi$ state was bound and, at the lower temperature, only the $J/\Psi$ and $\chi_c(1p)$ were bound.  In the charmonium sector, we found larger corrections to the adiabatic approximation, with the real-time evolution predicting stronger suppression in all cases.  This can be contrasted with bottomonium states for which we saw less $\Upsilon(1s)$ suppression using the real-time evolution at the lower initial temperature.  We also found that for both charmonia and bottomonia, the largest corrections to the adiabatic approximation were seen at low temperatures.  In order to draw firmer conclusions about the impact on bottomonium and charmonium observables such as $R_{AA}$, we need to include a realistic temperature profile and 3+1d viscous hydrodynamic evolution.

This study points to the need for developing efficient routines for solving the real-time Schr\"odinger evolution of heavy quarkonium states in the QGP that are coupled to realistic hydrodynamical backgrounds and a realistic complex-valued potential.  The need is more pressing for charmonium than it is for bottomonium, but in the end a unified framework which can describe both would be preferred.  Using the isotropic CKMS potential, it is technically feasible to fold together the real-time evolution with realistic 3+1d viscous hydrodynamics codes similar to what has been done in the case of the adiabatic approximation \cite{Strickland:2011mw,Strickland:2011aa,Strickland:2012cq,Krouppa:2015yoa,Krouppa:2016jcl,Krouppa:2017jlg}, however, the incorporation of the anisotropic non-equilibrium corrections \cite{Dumitru:2007hy,Burnier:2009yu,Dumitru:2009ni,Dumitru:2009fy,Nopoush:2017zbu} to the heavy quark potential will be technically challenging since the problem can no longer be reduced to a one-dimensional effective potential.

Another avenue for future improvement is to include a stochastic noise source similar to what was done in Ref.~\cite{Rothkopf:2013kya} to treat open quantum systems.  In general, one should use the trace- and positivity-preserving Lindblad form \cite{Akamatsu:2011se,Akamatsu:2012vt,Akamatsu:2014qsa,Katz:2015qja,Brambilla:2016wgg,Kajimoto:2017rel,Brambilla:2017zei,Blaizot:2017ypk,Blaizot:2018oev,Yao:2018nmy} for the evolution of the heavy quark system's reduced density matrix.  However, if the environment's relaxation time is short compared to the relaxation time of the system, then one can reduce the problem to solving a stochastic Schr\"odinger equation.  Two of us are currently working on extensions in this direction.

\acknowledgments{
M.S. and A.I. were supported by the U.S. Department of Energy, Office of Science, Office of Nuclear Physics under Award No. DE-SC0013470.
}

\appendix

\section{Numerical benchmarks}
\label{app:a}

In the body of the paper we used the split-step pseudospectral method for solving the real-time Schr\"odinger equation.  We would like to compare this with ``standard methods'' of solving the real-time Schr\"odinger equation in central potentials.  Below, we will refer to the split-step pseudospectral method applied to central potentials as the discrete sine transform (DST) method.  We will provide benchmark comparisons of the DST and Crank-Nicolson (CN) algorithms using both Mathematica and CUDA-based graphical processing unit (GPU) implementations.  For this purpose, we must first introduce the CN method.  In the CN method, one approximates the infinitesimal time evolution operator as
\be
e^{-i \hat{H} \Delta t} \simeq \frac{1 - \frac{i}{2} \Delta t \hat{H}}{1 + \frac{i}{2} \Delta t \hat{H}} \, ,
\ee
where the Hamiltonian has implicit time dependence through the time-dependence of the potential $V(t)$.  To proceed one moves the term in the denominator to left hand side after applying it to the wavefunction, i.e.
\be
\left(1 + \frac{i}{2} \Delta t \hat{H} \right) \psi(t+\Delta t) = \left(1 - \frac{i}{2} \Delta t \hat{H} \right)  \psi(t) \, .
\label{eq:CNupdate1}
\ee
To evolve the wave function one step forward in time, one applies the Hamilton on the right to obtain the ``half updated'' wave function $X \equiv \left(1 - \frac{i}{2} \Delta t \hat{H} \right)  \psi(t)$.  At this point, one must choose a discretization of the second derivative operator which appears in the kinetic energy term.  The standard choice is to use a three point formula (stencil) of the form $f''(x) = [ f(x-\Delta x) - 2 f(x) + f(x+\Delta x)]/(\Delta x)^2$.  As a result of this choice, the matrix on the left-hand-side of \eqref{eq:CNupdate1} is tridiagonal, $A \equiv \left(1 + \frac{i}{2} \Delta t \hat{H} \right)$.  In this way, one reduces the final step in the infinitesimal evolution of the wave function to solving a tridiagonal system of equations of the form
\be
A_{ij}  \psi_j(t+\Delta t) = X_i \, ,
\ee
where $i$ and $j$ index the spatial coordinates.  The tridiagonal system of equations can be solved using, e.g., the Thomas algorithm or optimized sparse linear solvers.

\begin{figure*}[t!]
\centerline{
\includegraphics[width=0.48\linewidth]{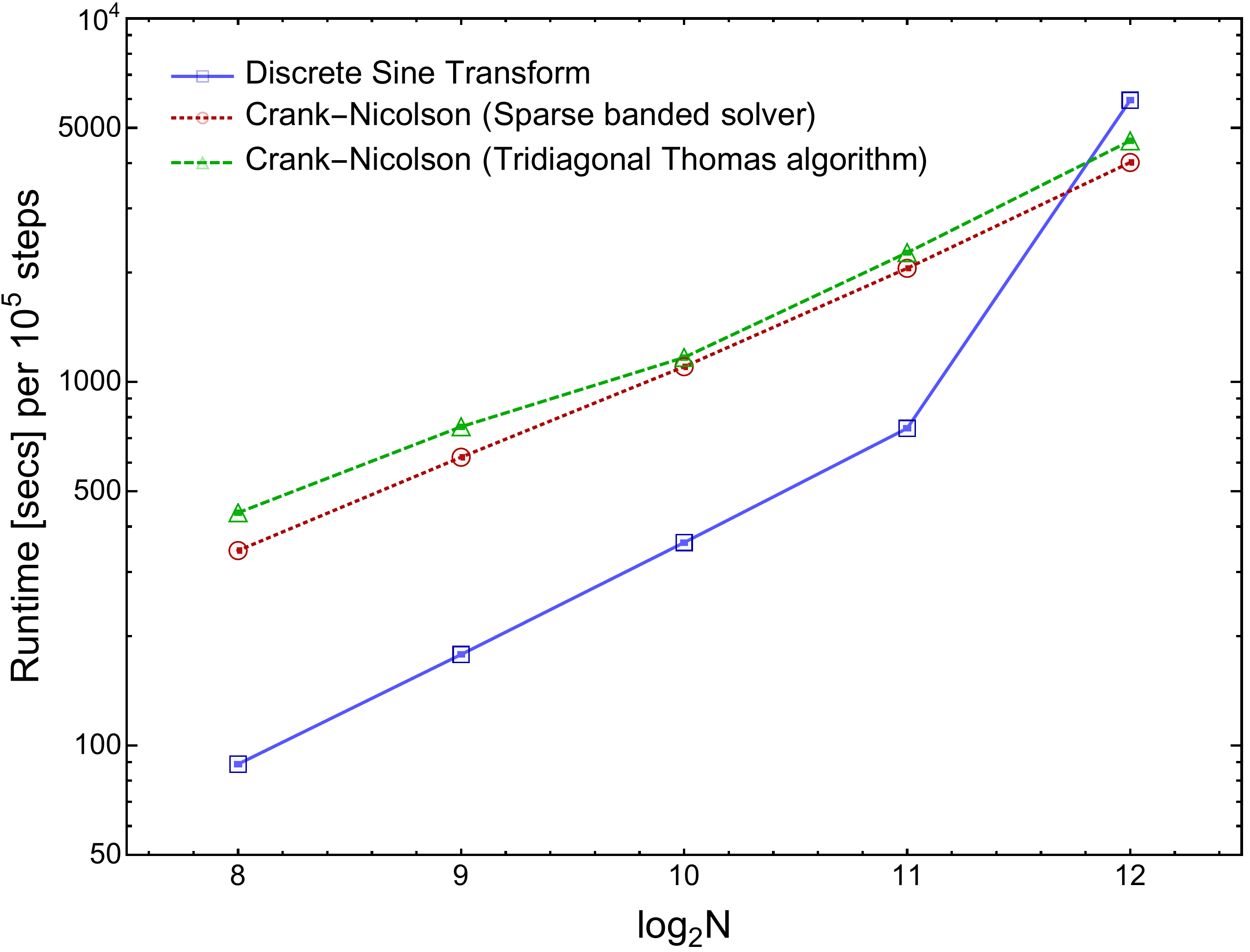}\hspace{5mm}
\includegraphics[width=0.48\linewidth]{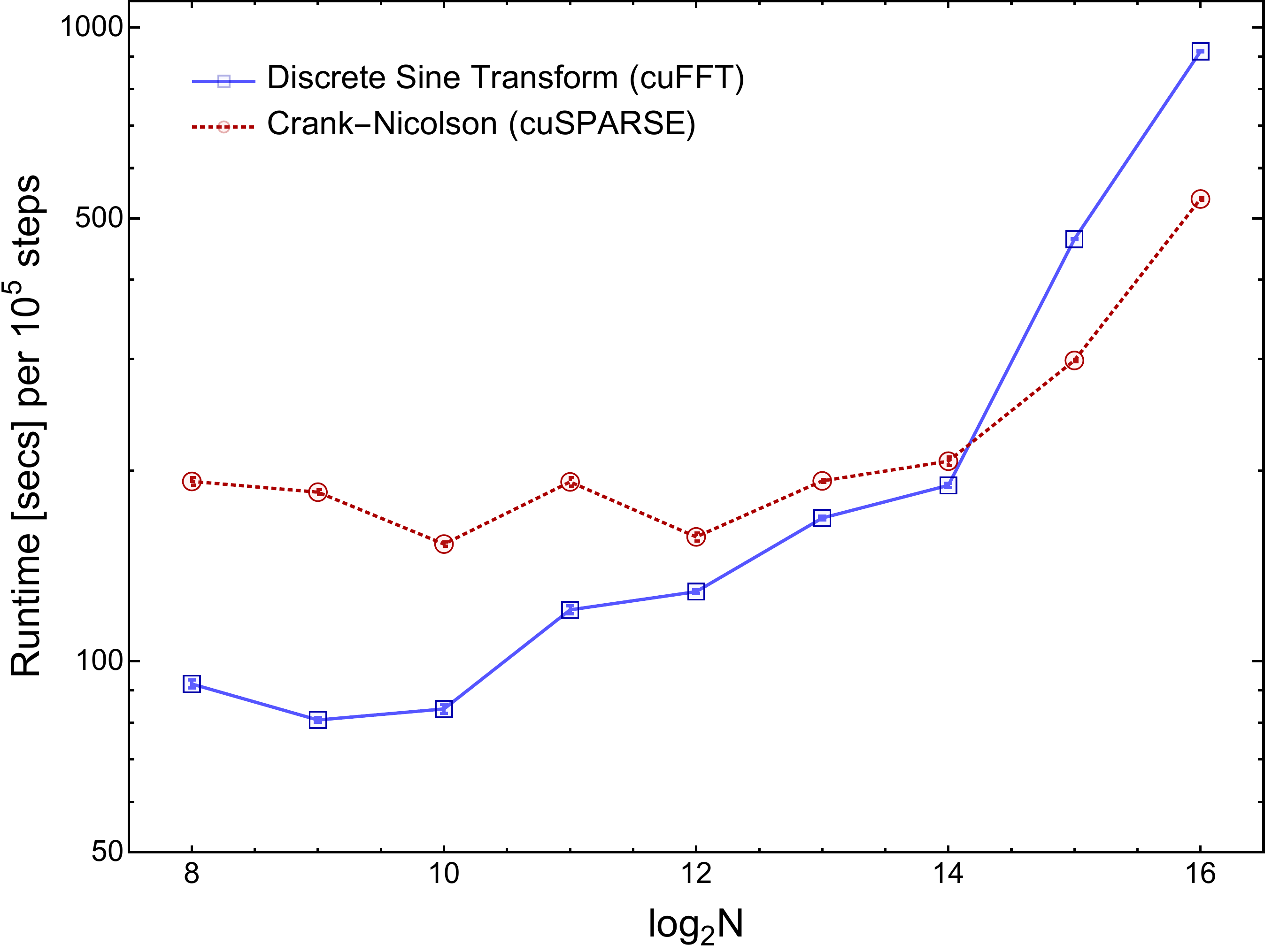}
}
\caption{Runtimes versus lattice size using Mathematica (left) and CUDA (right) implementations of the DST and CN algorithms}
\label{fig:benchmarks}
\end{figure*}

With this understanding, we can now start with the same initial wave function and time-dependent potential and perform the update using either the DST scheme described in the body of the paper or the CN scheme.  In all cases, both implementations give identical results up to numerical errors.  In Fig.~\ref{fig:benchmarks} we present our results for code runtimes in secs per $10^5$ steps.  The left panel shows the results obtained using Mathematica version 12 \cite{mathematica} implementation and the right panel shows the results obtained using the CUDA GPU graphics card programming language \cite{cuda}.  In the case of Mathematica (left panel), we provide comparisons between an algorithm using Mathematica's native DST, explicit implementation of the Thomas algorithm, and Mathematica's native sparse banded matrix solver.  The benchmarks were performed on a consumer level Mac Book Pro 2017 model with a 2.9 GHz Intel Core i7 processor.  As can be seen from the Mathematica benchmarks, the DST implementation is faster than the CN implementation for lattices size $N < 4096$.  In the case of the CUDA implemenations (right panel), we used a Nvidia Tesla K20m graphics card with 2496 cores.  For the DST implementation we made use of the cuFFT library for parallelized DST evaluation and, for the CN implementation, we use of the cuSPARSE parallelized tridiagonal matrix solver.  As can be seen from the right panel of Fig.~\ref{fig:benchmarks} the DST algorithm is faster for lattice sizes $N < 32768$.

We note that besides the improved performance observed, one major benefit of the DST algorithm is that the derivatives are computed using all points in the lattice, not just a three-point stencil.  As a result, the evolution obtained using the DST algorithm is more accurate than with the CN three-point derivative implementation. Of course, one could increase the accuracy of the CN algorithm by using a higher point number stencil, but this will come with an associated increase in compute time.  In general, the linear solver for a $M-$banded $N \times N$ matrix has a complexity of ${\cal O}(M^2 N)$.  As a result, one pays a steep price as the size of the stencil increases.  If one were to use a stencil of size $N$ which would provide the same accuracy for the derivatives as the DST, then the algorithmic complexity is ${\cal O}(N^3)$.  This should be contrasted with the DST algorithmic complexity which is ${\cal O}(N\log N)$.  Finally, we note that extending the DST algorithm to full 3d evolution is more straightforward than the corresponding CN implementation which requires, for example, the use of alternate direction implicit methods.

\bibliography{rtsol}

\end{document}